\documentclass[lettersize,journal]{IEEEtran}
\usepackage{amsmath,amsfonts,amssymb,amsthm}
\usepackage[ruled,linesnumbered,noend]{algorithm2e}
\SetArgSty{textnormal}
\usepackage{mathtools}
\usepackage[colorlinks,linkcolor=blue,citecolor=blue]{hyperref}
\usepackage{enumerate}
\usepackage{empheq}
\usepackage{array}
\usepackage{textcomp}
\usepackage{url}
\usepackage{verbatim}
\usepackage{graphicx}
\usepackage{cite}
\ifCLASSOPTIONcompsoc
\usepackage[caption=false,font=normalsize,labelfon
t=sf,textfont=sf]{subfig}
\else
\usepackage[caption=false,font=footnotesize]{subfig}
\fi
\def\BibTeX{{\rm B\kern-.05em{\sc i\kern-.025em b}\kern-.08em
    T\kern-.1667em\lower.7ex\hbox{E}\kern-.125emX}}

\usepackage{tikz}
\usetikzlibrary{arrows.meta}

\newtheorem{definition}{Definition}
\newtheorem{lemma}{Lemma}
\newtheorem{proposition}{Proposition}
\newtheorem{remark}{Remark}

\DeclareMathOperator{\dottriangledown}{\mathrel{\triangledown\hskip-0.4em\cdot\hskip0.15em}}
\DeclareMathOperator{\dotlozenge}{\mathrel{\lozenge\hskip-0.32em\cdot\hskip0.15em}}
\DeclareMathOperator{\vardottriangle}{\mathrel{\vartriangle\hskip-0.4em\cdot\hskip0.15em}}

\makeatletter
\newcommand{\labeltext}[3][]{%
    \@bsphack%
    \csname phantomsection\endcsname
    \def\tst{#1}%
    \def\labelmarkup{\emph}
    \def\refmarkup{}%
    \ifx\tst\empty\def\@currentlabel{\refmarkup{#2}}{\label{#3}}%
    \else\def\@currentlabel{\refmarkup{#1}}{\label{#3}}\fi%
    \@esphack%
    \labelmarkup{#2}
}
\makeatother

\newcommand{\Bin}{\{0,1\}}
\newcommand{\Frozen}{\mathcal{F}}
\newcommand{\mY}{\mathcal{Y}}
\newcommand{\mX}{\mathcal{X}}

\newcommand{\W}[2]{\mathbf{W}_{#1}^{(#2)}}
\newcommand{\V}[2]{\mathbf{V}_{#1}^{(#2)}}
\newcommand{\WABS}[2]{\mathbf{W}_{#1}^{(#2),\mathrm{ABS+}}}

\newcommand{\VABS}[2]{\mathbf{V}_{#1}^{(#2),\mathrm{ABS+}}}

\newcommand{\tildeW}[2]{\mathbf{\tilde W}_{#1}^{(#2)}}
\newcommand{\tildeV}[2]{\mathbf{\tilde V}_{#1}^{(#2)}}
\newcommand{\LLR}[2]{\mathbf{S}_{#1}^{(#2)}}
\newcommand{\LL}[2]{\mathbf{L}_{#1}^{(#2)}}
\newcommand{\RR}[2]{\mathbf{R}_{#1}^{(#2)}}
\newcommand{\MM}[2]{\mathbf{M}_{#1}^{(#2)}}

\newcommand{\PM}[1]{\mathbf{PM}_{#1}}
\newcommand{\GABS}[1]{G^\mathrm{ABS+}_{#1}}
\newcommand{\QABS}[1]{Q^\mathrm{ABS+}_{#1}}
\newcommand{\IS}[1]{\mathcal{I}_S^{(#1)}}
\newcommand{\IA}[1]{\mathcal{I}_A^{(#1)}}
\newcommand{\II}[1]{\mathcal{I}^{(#1)}}
\newcommand{\oria}{\triangledown}
\newcommand{\orib}{\lozenge}
\newcommand{\oric}{\vartriangle}
\newcommand{\swpa}{\blacktriangledown}
\newcommand{\swpb}{\blacklozenge}
\newcommand{\swpc}{\blacktriangle}
\newcommand{\adda}{{\dottriangledown}}
\newcommand{\addb}{{\dotlozenge}}
\newcommand{\addc}{{\vardottriangle}}

\newcommand{\Res}{\texttt{B}}
\newcommand{\Hlp}{\texttt{H}}
\newcommand{\LeftLLR}{\texttt{L}}
\newcommand{\RightLLR}{\texttt{R}}
\newcommand{\Mem}{\texttt{M}}

\DeclareMathOperator*{\argmax}{argmax}
\DeclareMathOperator{\sgn}{sgn}

\begin{document}

\title{Efficient LLR-Domain Decoding of ABS+ Polar Codes}

\author{Mikhail Chernikov, Peter Trifonov}



\maketitle

\begin{abstract}
ABS+ polar codes are a generalization of Arikan polar codes that provides much faster polarization. We present an LLR-domain version of the SCL decoder of ABS+ polar codes. Furthermore, we optimize the SCL algorithm in order to reduce the complexity of LLR computation. In comparison with classical polar codes, the proposed approach requires less number of arithmetic operations in the SCL decoder to obtain the same frame error rate (FER) at high-SNR region.
\end{abstract}


\section{Introduction}

\IEEEPARstart{P}{olar} codes \cite{arikan2009channel} achieve the capacity of a binary memoryless channel under successive cancellation (SC) decoding in the case of an infinite length. However, in the finite-length regime, the performance of SC is very poor. To overcome this issue one can use successive cancellation list (SCL) decoder \cite{tal2015list} which maintains a list of most likely prefixes of a message vector.  With a cyclic redundancy check (CRC), SCL decoder exhibits excellent error-correcting performance which is comparable to that of LDPC and turbo codes.

The complexity of the SCL decoder crucially depends on the list size \cite{xia2018highthroughput}. For a moderate blocklength the SCL decoder attains maximum likelihood (ML) even if the list size is pretty small. However, when the length is growing, the list size, required to get the near ML performance, increases exponentially \cite{mondelli2015scaling}. Several alternative decoding algorithms were recently proposed: sequential decoding \cite{miloslavskaya2014sequential}, SC flip and pertrubation decoding \cite{pillet2025successive}, sphere decoding \cite{hashemi2015list}, belief propagation decoding \cite{arikan2010polar} and permutation-based decoding \cite{kamenev2019permutation}. But, all these algorithms either are even more complicated or provide worse performance than SCL.

Recently discovered Adjacent-Bit-Swapped (ABS) \cite{li2023adjacent} and ABS+ polar codes \cite{li2024abs} entail faster polarization of virtual channels than Arikan polar codes. The encoding procedure is similar to encoding of classical polar codes, but to deepen the polarization, specific additional transforms of adjacent bits are performed. Due to the similarity to classical polar codes up to those transforms, ABS and ABS+ polar codes can be decoded by the decoder, analogous to the classical SC or SCL.

In this paper, we present the version of the SCL decoder for ABS+ polar codes that performs calculations over log-likelihood ratios (LLRs) instead of transition probabilities. This decoder uses only addition and comparison operations over real numbers, so it is well-suited for an efficient hardware implementation \cite{balatsoukasstimming2015llrbased}. Moreover, we show that lots of calculations in the originally presented SC algorithm are redundant. Indeed, at each phase it computes probabilities of two next message bits to be $0$ or $1$. However, it decodes only one bit dropping out the probabilities for the second one. Inspired by this idea, we provide a new version of the SC decoder (that can be easily generalized to SCL) not performing such unnecessary computations.

The proposed approach is compared against SCL decoder applied to classical polar codes. The numeric results show that for a fixed list size in the SCL decoder ABS+ codes demonstrate noticeably better performance. Besides, in many cases ABS+ decoder outperforms the standard one when the number of arithmetic operations in both decoders is fixed. On the other hand, the loss of the proposed decoder to the probability-domain decoder from \cite{li2024abs} is negligible.

The rest of this paper is organized as follows. In Section \ref{sect:background} the necessary background about classical and ABS+ polar codes is given. In Section \ref{sect:abs-llr} the LLR-domain decoder of ABS+ polar codes is proposed. Section \ref{sect:numeric-results} describes simulation results. Finally, takeaways and suggestions for a future work are discussed in Section \ref{sect:conclusions}.

\section{Background} \label{sect:background}

In this paper, we use the notation $x_i^j = (x_i, x_{i+1}, \dots, x_j)$. We  use $x_{i,o}^j$ and $x_{i,e}^j$ to denote the subsequences of $x_i^j$ consisting of entries with only odd and only even indices respectively. Concatenation of vectors $a$ and $b$ is denoted by $a.b$. For a channel $W$ the notation $W : \mX \to \mY$ defines $\mX$ and $\mY$ as input and output alphabets of $W$. The corresponding probability density function is denoted by $W(y|x),\ x \in \mX,\ y \in \mY$. Kronecker product of matrices $A$ and $B$ is denoted by $A \otimes B$, $A^{\otimes m}$ corresponds to the $m$-th Kronecker power of $A$. All operations over binary values are performed modulo two.

\subsection{Classical polar codes} \label{subsect:polar-codes}

A polar code is a binary linear block code whose codewords are vectors $u_1^n F^{\otimes m}$, where $n=2^m$ is the length, $F = \left(\begin{smallmatrix} 1 & 0 \\ 1 & 1 \end{smallmatrix}\right)$ and $u_1^n \in \Bin^n$ has zeros at all the positions from the predefined set $\Frozen \subseteq \{1,2,\dots,n\}$ called frozen set. In what follows the vector $u_1^n$ is referred to as an input vector and $n=2^m$ is reserved for code length.

Multiplying by $G_m = F^{\otimes m}$ results in the channel polarization effect \cite{arikan2009channel}. Consider a binary descrete memoryless channel (B-DMC) $W : \Bin \to \mY$ and let us transmit entries of the vector $u_1^nG_m$ through $n$ independent copies of $W$. This gives rise to $n$ virtual bit subchannels $\W{i}{m} : \Bin \to \mY^n \times \Bin^{i-1} ,\ 1 \leq i \leq n$ with transition probabilities
\begin{equation}
\label{eq:1bit-subchannels}
\begin{split}
        \W{i}{m}&(y_1^n,u_1^{i-1}|u_i) = \frac{1}{2^{n-1}} \sum_{u_{i+1}^n \in \Bin^{n-i}} \W{}{m}(y_1^n | u_1^n),
\end{split}
\end{equation}
where $\W{}{m}(y_1^n | u_1^n) = \prod_{j=1}^n W(y_j|(u_1^nG_m)_j)$ is the probability to receive the vector $y_1^n$ if $u_1^nG_m$ has been transmitted.

When $n$ grows to infinity, every subchannel becomes either absolutely noisy or absolutely noiseless. In order to transmit information in the most efficient way, indices of the least reliable bit subchannels are included to $\Frozen$. The remaining positions of the message vector carry payload.

\subsection{Successive cancellation decoding} \label{subsect:arikan-sc-decoding}

The SC decoding algorithm successively estimates bits of the input vector. To get the $i$-th bit, it calculates the transition probabilities of the corresponding virtual subchannel with respect to the received vector $y_1^n$ and previously decoded message bits $\hat u_1^{i-1}$. Then, it makes the hard decision:
\begin{equation}
        \hat u_i = \begin{cases}
                \argmax\limits_{u_i \in \Bin} \W{i}{m}(y_1^n, \hat u_1^{i-1} | u_i), & i \not\in \Frozen,\\
                0, & \text{otherwise}.
        \end{cases}
\end{equation}
After all input bits are decoded, the corresponding codeword $\hat c_1^n = \hat u_1^n G_m$ is returned.

The procedure of calculation of these transition probabilities can be seen as a binary tree traversal (for example, see \cite[Fig. 1]{hashemi2017fast}). In fact, there is a recursive relation between virtual subchannels. In the polarization scheme depicted in Fig. \ref{fig:1bit-subchannels}, two copies of B-DMC $W: \Bin \to \mY$ are transformed into virtual channels $W^- : \Bin \to \mY$ and $W^+: \Bin \to \mY \times \Bin$:
\begin{subequations}
        \begin{align}
                &W^-(y_1,y_2|u_1) = \frac{1}{2}\sum_{u_2 \in \Bin} W(y_1|u_1+u_2)W(y_2|u_2),\\
                &W^+(y_1,y_2,u_1|u_2) = \frac{1}{2} W(y_1|u_1+u_2)W(y_2|u_2).
        \end{align}
\end{subequations}
Then, for every $1 \leq i \leq 2^{m-1}$ one has
\begin{equation}
\label{eq:1bit-subchannel-evolution}
        \W{2i-1}{m} = (\W{i}{m-1})^-,\ \W{2i}{m} = (\W{i}{m-1})^+.
\end{equation}

To make the computations hardware-friendly, the transition probabilities can be replaced with approximate LLRs for the corresponding virtual subchannels \cite{balatsoukasstimming2015llrbased}, \cite{trifonov2018score}. Let us define the approximate LLRs as follows:
\begin{equation}
        \label{eq:arikan-llr}
        \LLR{i}{m}(y_1^n,u_1^{i-1}) = \ln\frac{\tildeW{i}{m}(y_1^n, u_1^{i-1} | 0)}{\tildeW{i}{m}(y_1^n, u_1^{i-1} | 1)},
\end{equation}
where $1 \leq i \leq n$ and
\begin{equation}
        \label{eq:arikan-approx-probs}
        \tildeW{i}{m}(y_1^n, u_1^{i-1} | u_i) = \max\limits_{u_{i+1}^n \in \Bin^{n-i}} \W{}{m}(y_1^n | u_1^n)
\end{equation}
is the approximation of the transition probability of $\W{i}{m}$. For the computation of the approximate LLRs, the decoder applies the following expressions recursively:
\begin{subequations}
\begin{align}
        &\LLR{2i-1}{m}(y_1^n,u_1^{2i-2}) = f_-(a,b) \triangleq \sgn(a)\sgn(b)\min\{|a|,|b|\},\\
        &\LLR{2i}{m}(y_1^n,u_1^{2i-1}) = f_+(a,b,u_{2i-1}) \triangleq (-1)^{u_{2i-1}}a + b,
\end{align}
\end{subequations}
where $a = \LLR{i}{m-1}(y_{1,o}^n, u_{1,o}^{2i-2} + u_{1,e}^{2i-2}),\ b = \LLR{i}{m-1}(y_{1,e}^n, u_{1,e}^{2i-2})$. The initial values are $\LLR{1}{0}(y_\beta) = \ln\dfrac{W(y_\beta|0)}{W(y_\beta|1)},\ 1 \leq \beta \leq n$. The hard decisions are obtained as
\begin{equation}
        \hat u_i = \begin{cases}
                1, & \LLR{i}{m}(y_1^n, \hat u_1^{i-1}) < 0\ \text{and}\ i \not\in \Frozen, \\
                0, & \text{otherwise}.
        \end{cases}
\end{equation}

\begin{figure}
        \centering
        \begin{tikzpicture}

\tikzstyle{every node}=[font=\normalfont]

\node at (0.5,1.5) {$u_1$};
\node at (0.5,0) {$u_2$};

\node at (2.5,1.5) {$\bigoplus$};

\draw [-{Stealth[length=2mm]}] (0.75,1.5) -- (2.325,1.5);
\draw [-{Stealth[length=2mm]}] (2.625,1.5) -- (4,1.5);

\draw [-{Stealth[length=2mm]}] (0.75,0) -- (4,0);
\draw [-{Stealth[length=2mm]}] (2.5,0) -- (2.5,1.325);

\node at (4.25,1.5) {$W$};
\node at (4.25,0) {$W$};

\draw  (4,2) rectangle (4.5,1);
\draw  (4,0.5) rectangle (4.5,-0.5);

\draw [-{Stealth[length=2mm]}] (4.5,1.5) -- (5,1.5);
\draw [-{Stealth[length=2mm]}] (4.5,0) -- (5,0);

\node at (5.25,1.5) {$y_1$};
\node at (5.25,0) {$y_2$};

\end{tikzpicture}
        \caption{Polarization of a channel $W: \Bin \to \mY$.}
        \label{fig:1bit-subchannels}
\end{figure}

\subsection{Successive cancellation list decoding}

Although SC decoding allows polar codes to achieve channel capacity for $n \to \infty$, for finite $n$ its performance is far away from the maximum likelihood (ML). It can be improved by using the SCL algorithm \cite{tal2015list}. We are to review its min-sum version \cite{balatsoukasstimming2015llrbased}.

The SCL algorithm maintains up to $L$ possible prefixes of the input vector which are referred to as paths. Here $L$ is the predefined parameter called list size. The decoder recursively computes the values $\LLR{i}{m}(y_1^n, \hat u_1^{i-1})$ for each path (so, here $\hat u_1^{i-1}$ denotes decoded bits of some considered path) in the same way as in SC. If $i \in \Frozen$, all paths are extended with $\hat u_i = 0$. Otherwise, the decoder splits every path into two continuations with $\hat u_i = 0$ and $\hat u_i = 1$. After that, some least likely paths are eliminated in order to get at most $L$ paths present in the list.

To determine path likelihood after the $i$-th input bit has been processed, the decoder computes for each active path the number $\PM{i}$ called path metric (PM). Greater metric corresponds to a less reliable path. Initially there are only one active path whose metric is zero. The PMs are updated as
\begin{equation}
\label{eq:path-metrics}
    \PM{i} = \begin{cases}
            \PM{i-1}, & \hat u_i = \dfrac{1}{2}(1-\sgn(\alpha)), \\
            \PM{i-1} + |\alpha|, & \text{otherwise},
    \end{cases}
\end{equation}
where $\alpha = \LLR{i}{m}(y_1^n, \hat u_1^{i-1})$ and $\hat u_1^{i-1}$ is the path prefix continued with $\hat u_i$. Saying less formally, each path is penalized when its hard decision does not match the sign of the LLR. As the result, the decoder returns the codeword corresponding to the most reliable path after all $n$ input bits are decoded.

\subsection{ABS+ polar codes} \label{subsect:abs-polar-codes}

For $n \to \infty$ the capacity of every virtual channel in \eqref{eq:1bit-subchannels} tends either to $0$ or to $1$ \cite{arikan2009channel}, but for a moderate code length there are many subchannels that are not fully polarized, i.e. whose capacities belong to the interval $(\delta,1-\delta)$ for some sufficiently large $\delta \in (0, \frac{1}{2})$. This may result in a poor performance of the SC decoder or the SCL decoder with small list size. To deepen the polarization in the finite-length regime, authors in \cite{li2023adjacent} and \cite{li2024abs} have suggested to perform specific transforms over adjacent bits while encoding. This gives rise to virtual subchannels that achieve the polarized state, when $n$ is growing, provably faster than the channels in \eqref{eq:1bit-subchannels}. Therefore, polar codes based on this idea (called ABS+ polar codes \cite{li2024abs}) provide substantially better SC/SCL performance with respect to the classical polar codes.

The polarization matrix of a classical polar code is defined as $G_m = G_{m-1} \otimes F$. When it comes to an ABS+ polar code, we have matrix $\GABS{m} = \QABS{m}(\GABS{m-1} \otimes F)$ with the initial condition $\GABS{1} = F$, where $\QABS{m}$ is the $n \times n$ matrix, corresponding to the mentioned transforms of adjacent bits. Namely, $\QABS{m}$ is defined by two sets $\IS{m}, \IA{m} \subseteq \{1,2,\dots,n-1\}$ and is represented as
\begin{equation}
\label{eq:qabs-def}
        \QABS{m} = \left( \prod_{i \in \IS{m}} S_n{(i)} \right) \left( \prod_{i \in \IA{m}} A_n{(i)} \right),
\end{equation}
where $S_n{(i)}$ and $A_n{(i)}$ are $n \times n$ binary matrices such that for every $x_1^n \in \Bin^n$ one has
\begin{enumerate}
        \item $x_1^nS_n{(i)} = (x_1, \dots, x_{i-1}, x_{i+1}, x_i, x_{i+2}, \dots, x_n),$
        \item $x_1^nA_n{(i)} = (x_1, \dots, x_{i-1}, x_i + x_{i+1}, x_{i+1}, \dots, x_n).$
\end{enumerate}
In the first case we have entries $x_i$ and $x_{i+1}$ swapped and in the second case we have $x_{i+1}$ added to $x_i$ while the other elements of $x_1^n$ are kept unchanged. Furthermore, the sets $\IS{m}$ and $\IA{m}$ must satisfy two following constraints:
\begin{enumerate}[(i)]
        \item \label{item:cond-1} $\IS{m} \cap \IA{m} = \emptyset$.
        \item \label{item:cond-2} Let $\II{m} \triangleq \IS{m} \cup \IA{m} = \{i_1,i_2,\dots,i_s\}$. Then $$i_1 + 4 \leq i_2,\ i_2 + 4 \leq i_3,\ \dots,\ i_{s-1} + 4 \leq i_s.$$
\end{enumerate}
Of course, these conditions are also held for all $\QABS{\lambda},\ 2 \leq \lambda < m$.
So, an ABS+ polar code is defined by a frozen set $\Frozen \subseteq \{1,2,\dots,n\}$ and sets $\{\IS{\lambda}, \IA{\lambda}\}_{\lambda=2}^m$ and the codewords have the form $u_1^n\GABS{m}$ with $u_1^n \in \Bin^n$ such that $u_i = 0$ whenever $i \in \Frozen$.

The matrix $\QABS{m}$ is chosen in the way that leads to fastest polarization. In \cite{li2024abs} the authors formulate the corresponding optimization problem and propose an explicit algorithm solving it. In that work, it is also noticed that for the purpose of deepening polarization by using transforms of adjacent bits, it is sufficient to take into account only transforms $S_n{(i)}$ and $A_n{(i)}$ for all $1 \leq i \leq n-1$. Moreover, a transformation over the $i$-th and $(i+1)$-th bits may improve polarization only if $i$ is even, so all elements of $\II{m}$ must be even numbers.

Let us describe encoding of ABS+ polar codes more precisely. It can be viewed as passing an input vector $u_1^n$ through the encoding circuit \cite{arikan2010polar} of $m$ layers. Initially, let $u_{i,1}^{(m)} = u_i,\ 1 \leq i \leq n$. Next, for $0 \leq \lambda < m,\ 1 \leq i \leq 2^\lambda,\ N=2^{m-\lambda}$ consider $u_i^{(\lambda)} = (u_{i,1}^{(\lambda)}, \dots, u_{i,N}^{(\lambda)})$ such that
\begin{subequations}
\label{eq:abs-component-vectors}
\begin{empheq}[left={u_i^{(\lambda)} =\empheqlbrace}]{alignat=2}
        & \begin{aligned}
                        (u_{2i-1}^{(\lambda+1)} +& u_{2i}^{(\lambda+1)}\ .\ u_{2i}^{(\lambda+1)}),\\ & 2i \not\in \II{\lambda + 1},\ 2(i-1) \not\in \IS{\lambda+1}
                \end{aligned} \label{eq:abs-component-vectors-1}\\
        & \begin{aligned}
                (u_{2i-2}^{(\lambda+1)} + u_{2i}^{(\lambda+1)}\ &.\ u_{2i}^{(\lambda+1)}), \\ &2(i-1) \in \IS{\lambda+1},
        \end{aligned} \label{eq:abs-component-vectors-2}\\
        & (u_{2i-1}^{(\lambda+1)} + u_{2i+1}^{(\lambda+1)}\ . \ u_{2i+1}^{(\lambda+1)}),\ 2i \in \IS{\lambda+1}, \label{eq:abs-component-vectors-3}
                \\
        & \begin{aligned}
                        (u_{2i-1}^{(\lambda+1)} + & u_{2i}^{(\lambda+1)} + u_{2i+1}^{(\lambda+1)}\ .\\ & u_{2i}^{(\lambda+1)} + u_{2i+1}^{(\lambda+1)}),\ 2i \in \IA{\lambda+1}.
                \end{aligned} \label{eq:abs-component-vectors-4}        
\end{empheq}
\end{subequations}
Then, the resulting codeword is $u_1^{(0)}$. In fact, the case \eqref{eq:abs-component-vectors-1} corresponds to standard polarizing transform $(a,b) \mapsto (a+b.b)$ applied to $u_{2i-1}^{(\lambda+1)}$ and $u_{2i}^{(\lambda+1)}$. When $2i \in \IS{\lambda+1}$, we swap the vectors $u_{2i}^{(\lambda+1)}$ and $u_{2i+1}^{(\lambda+1)}$  (see \eqref{eq:abs-component-vectors-2}-\eqref{eq:abs-component-vectors-3}). Finally, if $2i \in \IA{\lambda+1}$, $u_{2i+1}^{(\lambda+1)}$ is added to $u_{2i}^{(\lambda+1)}$ (see \eqref{eq:abs-component-vectors-4}). An example of encoding circuit for an ABS+ polar code is given in Fig. \ref{fig:abs-factor-graph}.

\begin{figure}
        \centering
        \begin{tikzpicture}[
        scale=0.75,
        -{Stealth[length=1.5mm]}
]

\tikzstyle{every node}=[font=\fontsize{2.75mm}{2mm}\selectfont]
\node at (-2,2) {$0$};
\node at (-2,1) {$0$};
\node at (-2,0) {$0$};
\node at (-2,-1) {$0$};
\node at (-2,-2) {$u_5$};
\node at (-2,-3) {$u_6$};
\node at (-2,-4) {$u_7$};
\node at (-2,-5) {$u_8$};

\node at (0,2) {$\bigoplus$};
\node at (0,0) {$\bigoplus$};
\node at (0,-2) {$\bigoplus$};
\node at (0,-4) {$\bigoplus$};

\draw (-1.75,2) -- (-0.175,2);
\draw (-1.75,0) -- (-0.175,0);
\draw [orange] (-1.75,-2) -- (-1,-1) -- (0.5,-1);
\draw (-1.75,-4) -- (-0.175,-4);

\draw (0.175,2) -- (0.5,2);
\draw (0.175,0) -- (0.5,0);
\draw (0.175,-2) -- (0.5,-2);
\draw (0.175,-4) -- (0.5,-4);

\draw (-1.75,1) -- (0.5,1);
\draw [orange] (-1.75,-1) -- (-1,-2) -- (-0.175,-2);
\draw (-1.75,-3) -- (0.5,-3);
\draw (-1.75,-5) -- (0.5,-5);

\draw (0,1) -- (0,1.825);
\draw (0,-1) -- (0,-0.175);
\draw (0,-3) -- (0,-2.175);
\draw (0,-5) -- (0,-4.175);

\node at (1,1.5) {$u_1^{(2)}$};
\node at (1,-0.5) {$u_2^{(2)}$};
\node at (1,-2.5) {$u_3^{(2)}$};
\node at (1,-4.5) {$u_4^{(2)}$};

\draw[dashed, red] (0.5,2.25) rectangle (1.5,0.75);
\draw[dashed, red] (0.5,0.25) rectangle (1.5,-1.25);
\draw[dashed, red] (0.5,-1.75) rectangle (1.5,-3.25);
\draw[dashed, red] (0.5,-3.75) rectangle (1.5,-5.25);

\node at (3,2) {$\bigoplus$};
\node at (3.5,1) {$\bigoplus$};
\node at (3,-2) {$\bigoplus$};
\node at (3.5,-3) {$\bigoplus$};

\node[orange] at (2,0) {$\bigoplus$};
\node[orange] at (2.5,-1) {$\bigoplus$};

\draw (1.5,2) -- (2.825,2);
\draw (1.5,1) -- (3.325,1);
\draw (1.5,-2) -- (2.825,-2);
\draw (1.5,-3) -- (3.325,-3);

\draw (1.5,2) -- (4,2);
\draw (1.5,1) -- (4,1);
\draw (1.5,-2) -- (4,-2);
\draw (1.5,-3) -- (4,-3);

\draw (1.5,0) -- (1.825,0);
\draw (1.5,-1) -- (2.325,-1);
\draw (1.5,-4) -- (4,-4);
\draw (1.5,-5) -- (4,-5);

\draw (2.175,0) -- (4,0);
\draw (2.675,-1) -- (4,-1);

\draw (3,0) -- (3,1.875);
\draw (3,-4) -- (3,-2.175);
\draw (3.5,-1) -- (3.5,0.825);
\draw (3.5,-5) -- (3.5,-3.175);

\draw [orange] (2,-2) -- (2,-0.175);
\draw [orange] (2.5,-3) -- (2.5,-1.175);

\node at (4.5,0.5) {$u_1^{(1)}$};
\node at (4.5,-3.5) {$u_2^{(1)}$};
\draw[dashed, red] (4,2.25) rectangle (5,-1.25);
\draw[dashed, red] (4,-1.75) rectangle (5,-5.25);

\node at (5.5,2) {$\bigoplus$};
\node at (6,1) {$\bigoplus$};
\node at (6.5,0) {$\bigoplus$};
\node at (7,-1) {$\bigoplus$};

\draw (5,-2) -- (7.5,-2);
\draw (5,-3) -- (7.5,-3);
\draw (5,-4) -- (7.5,-4);
\draw (5,-5) -- (7.5,-5);

\draw (5.5,-2) -- (5.5,1.825);
\draw (6,-3) -- (6,0.825);
\draw (6.5,-4) -- (6.5,-0.175);
\draw (7,-5) -- (7,-1.175);

\draw (5,2) -- (5.325,2);
\draw (5,1) -- (5.825,1);
\draw (5,0) -- (6.325,0);
\draw (5,-1) -- (6.825,-1);

\draw (5,2) -- (7.5,2);
\draw (5,1) -- (7.5,1);
\draw (5,0) -- (7.5,0);
\draw (5,-1) -- (7.5,-1);

\node at (8,-1.5) {$u_1^{(0)}$};
\draw[dashed, red] (7.5,2.25) rectangle (8.5,-5.25);

\end{tikzpicture}
        \caption{Encoding of $(8,4)$ ABS+ polar code with $\Frozen = \{1,2,3,4\}$, $\II{3} = \IS{3} = \{4\},\ \II{2} = \IA{2} = \{2\}$. Adjacent bit transforms are highlighted.}
        \label{fig:abs-factor-graph}
\end{figure}

\subsection{SC decoding of ABS+ polar codes in probability domain}

The SC decoder of ABS+ polar codes also successively decodes input bits considering the virtual subchannels. More precisely, similarly to \eqref{eq:1bit-subchannels}, one can introduce the channels
\begin{equation}
\label{eq:abs-subchannels}
\begin{split}
        \WABS{i}{m}&(y_1^n, u_1^{i-1} | u_i) = \\ = \frac{1}{2^{n-1}} & \sum_{u_{i+1}^n \in \Bin^{n-i}} \WABS{}{m}(y_1^n | u_1^n).
\end{split}
\end{equation}
with $\WABS{}{m}(y_1^n|u_1^n) = \prod_{j=1}^n W(y_j|(u_1^n\GABS{m})_j)$. In what follows we omit the upper subscript "$\mathrm{ABS+}$" in channel designations, always considering virtual channels stemming from the polarizing transform $u_1^n \mapsto u_1^n\GABS{m}$. For instance, we use $\W{i}{m} \equiv \WABS{i}{m}$.

Unfortunately, between the subchannels \eqref{eq:abs-subchannels} there does not exist an explicit recursive relation (analogous to \eqref{eq:1bit-subchannel-evolution}) that could be used by the decoder. Instead, one can introduce double-bit-input (DBI) virtual subchannels $\V{i}{m} : \Bin^2 \to \mY^n \times \Bin^{i-1},\ 1 \leq i < n$ whose input represents two adjacent bits (the $i$-th and the $(i+1)$-th one) of an input vector and the output is the preceding prefix together with received values from the copies of $W$. Lemma 3 from \cite{li2024abs} establishes a recursive relation between these channels for computing the transition probabilities.

Before we explore the decoding algorithm, let us introduce some notations. We use $y_1^n \in \mY^n$ to denote a received vector, $\hat u_i$ is an estimate of the $i$-th input bit and $\hat u_i^{(\lambda)}$ is an estimate of $u_i^{(\lambda)}$ from \eqref{eq:abs-component-vectors}. We also use the shortcut
\begin{equation*}
        \hat x_{i,\beta}^{(\lambda)} = (y_\beta, y_{\beta + N}, \dots, y_{\beta + (2^\lambda-1)N}, \hat u_{1,\beta}^{(\lambda)}, \hat u_{2,\beta}^{(\lambda)}, \dots, \hat u_{i-1,\beta}^{(\lambda)})
\end{equation*}
with $0 \leq \lambda \leq m,\ 1 \leq i \leq 2^{\lambda},\ 1 \leq \beta \leq N = 2^{m-\lambda}$.

Note that for $1 \leq \lambda \leq m$ and $1 \leq i \leq 2^\lambda$ such that $2i \in \II{\lambda+1}$, in order to get $u_i^{(\lambda)}$, the vectors $u_{2i-1}^{(\lambda+1)},\ u_{2i}^{(\lambda+1)},\ u_{2i+1}^{(\lambda+1)}$ must be decoded (see \eqref{eq:abs-component-vectors-3} and \eqref{eq:abs-component-vectors-4}). Moreover, according to \cite[Lemma 3]{li2024abs}, in these cases the calculation of the transition probabilities of $\V{2i+1}{\lambda+1}$ is based on the transition probabilities of $\V{i}{\lambda}$. The SC decoder of classical polar codes, in contrast, recursively processes only the $(2i-1)$-th and the $(2i)$-th channels to decode $u_i^{(\lambda)}$. Another point to mention is that the virtual subchannels $\V{i}{\lambda}$ are defined only for $1 \leq i \leq 2^\lambda-1$. So, $u_{2^\lambda-1}^{(\lambda)}$ is being decoded together with $u_{2^\lambda}^{(\lambda)}$ while processing $\V{2^\lambda-1}{\lambda}$. The last remark is that if $2(i-1) \in \II{\lambda+1}$, $\hat u_i^{(\lambda)} = (v + \hat u_{2i}^{(\lambda+1)} . \hat u_{2i}^{(\lambda+1)})$, where $v$ is already decoded before processing $\V{i}{\lambda}$. Hence, $v$ is stored in a specific buffer after it has been decoded and is copied when it is necessary to calculate $\hat u_i^{(\lambda)}$.

\begin{figure}
        \centering
        \begin{tikzpicture}[-{Stealth[length=2mm]}]

\node (u11) at (-2,-1) {$(1,1)$};

\node (u21) at (-3,-2) {$(2,1)$};
\node (u22) at (-1,-2) {$(2,2)$};
\node (u23) at (1,-2) {$(2,3)$};

\node (u31) at (-3.5,-3) {$(3,1)$};
\node (u32) at (-2.5,-3) {$(3,2)$};
\node (u33) at (-1.5,-3) {$(3,3)$};
\node (u34) at (-0.5,-3) {$(3,4)$};
\node (u35) at (0.5,-3) {$(3,5)$};
\node (u36) at (1.5,-3) {$(3,6)$};
\node (u37) at (2.5,-3) {$(3,7)$};

\draw[red] (u11) edge (u21);
\draw[red] (u11) edge (u22);
\draw[red] (u11) edge (u23);

\draw[red] (u21) edge (u31);
\draw[red] (u21) edge (u32);
\draw[red] (u21) edge (u33);
\draw (u22) edge (u34);
\draw (u23) edge (u35);
\draw (u23) edge (u36);
\draw (u23) edge (u37);
        
\end{tikzpicture}
        \caption{The tree of the recursion in the SC algorithm applied to an ABS+ polar code with $n=8,\ \II{2} = \{2\},\ \II{3} = \{2\}$.  We use a black edge from $(\lambda,i)$ to $(\lambda+1,i')$ if $2i \not\in \II{\lambda+1}$. Otherwise the edge is red.}
        \label{fig:abs-sc-tree-1}
\end{figure}

The decoding consists of recursive computation of transition probabilities for the virtual DBI channels, making hard decisions and obtaining the resulting codeword by application of \eqref{eq:abs-component-vectors} to estimated input bits. The recursion can be seen as a traversal of the tree which represents dependencies between virtual DBI channels and intermediate vectors in \eqref{eq:abs-component-vectors}. An example of such tree is depicted in Fig. \ref{fig:abs-sc-tree-1}. Every node is labeled by a pair of integers $(\lambda,i)$ such that $1 \leq \lambda \leq m,\ 1 \leq i \leq 2^\lambda-1$. Here $\lambda$ is referred to as number of layer, $i$ is referred to as number of phase. The node $(1,1)$ is the tree root. Leafs are the nodes from the $m$-th layer. All edges have the form $(\lambda,i) \to (\lambda+1,j)$, where $j \in \{2i-1,2i,2i+1\}$. The edge $(\lambda,i) \to (\lambda+1,j)$ indicates that $\V{j}{\lambda+1}$ can be obtained by applying one of the transforms $(-)^\oria, (-)^\orib, (-)^\oric, (-)^\swpa, (-)^\swpb, (-)^\swpc, (-)^\adda, (-)^\addb, (-)^\addc$ (see \cite{li2024abs}) to $\V{i}{\lambda}$. As the input for processing of the node $(\lambda,i)$ the algorithm receives the array of values $\V{i}{\lambda}(\hat x_{i,\beta}^{(\lambda)}|a,b),\ 1 \leq \beta \leq 2^{m-\lambda},\ a,b \in \Bin$. As the output, it returns $\hat u_i^{(\lambda)}$ and, if $i = 2^\lambda-1$, $\hat u_{i+1}^{(\lambda)}$ is also returned. To process the leaf node $(m,i)$, the algorithm makes a hard decision on $u_i$ using
\begin{align*}
        \W{i}{m}(y_1^n, u_1^{i-1} | u_i) &= \frac{1}{2}\sum_{u_{i+1} \in \Bin} \V{i}{m}(y_1^n, u_1^{i-1} | u_i, u_{i+1}),\\
        \W{i+1}{m}(y_1^n, u_1^i | u_{i+1}) &= \frac{1}{2}\V{i}{m}(y_1^n, u_1^{i-1} | u_i, u_{i+1}). \label{eq:last-bit-prob}
\end{align*}
If $i=n-1$, $u_{i+1}$ is also decoded. Non-leaf nodes are being processed by recursive visiting of the child nodes in the ascending order of the phases. The transition probabilities for children of $(\lambda,i)$ are being computed with respect to the known values $\V{i}{\lambda}(\hat x_{i,\beta}^{(\lambda)}|a,b), 1 \leq \beta \leq 2^{m-\lambda},\ a,b \in \Bin$ and the previously decoded bits on the $(\lambda+1)$-th layer. The resulting vector $\hat u_i^{(\lambda)}$ is calculated with respect to the results of children processing (and the value stored at the $(i-1)$-th phase if $2(i-1) \in \II{\lambda+1}$). The initial probabilities for processing of the root node are
\begin{equation*}
        \V{1}{1}(y_\beta,y_{\beta+n/2}|a,b) = W(y_\beta|a+b)W(y_{\beta+n/2}|b).
\end{equation*}
The resulting codeword can be easily obtained from the vectors $\hat u_1^{(1)},\ \hat u_2^{(1)}$ representing the output after processing of the root.

\section{LLR-Domain Computations for Decoding of ABS+ Polar Codes} \label{sect:abs-llr}

As established in \cite{balatsoukasstimming2015llrbased}, a hardware implementation of SC-like decoders can be significantly simplified by using LLRs instead of transition probabilities. In the case of ABS+ polar codes calculations of these probabilities are even more complicated than for classical polar codes. Hence, it is necessary to adapt the decoder to LLR domain.

We start from the definitions of approximate transition probabilities for of virtual DBI channels from \eqref{eq:abs-subchannels} in Section \ref{subsect:llr-evolution}. Then we establish a recursive relation between the approximate probabilities and present the formulae for evaluation of the LLRs. In Section \ref{subsect:reusing} and Section \ref{subsect:unused-llrs} we analyse these formulae and suggest methods of reduced complexity computations. In Section \ref{subsect:llr-algorithm} we formulate the efficient SC decoding algorithm in the LLR domain. Generalization of this SC decoder to SCL is briefly discussed in Section \ref{subsect:llr-scl}.

\subsection{Approximate LLRs for the virtual DBI subchannels} \label{subsect:llr-evolution}

Similarly to \eqref{eq:arikan-approx-probs}, we define the approximate probabilities for the virtual bit subchannels
\begin{equation}
        \tildeW{i}{m}(y_1^n, u_1^{i-1} | u_i) = \max_{u_{i+1}^n \in \Bin^{n-i}} \W{}{m}(y_1^n | u_1^n)
\end{equation}
with $1 \leq i \leq n$ and also for the defined earlier DBI subchannels:
\begin{equation}
\label{eq:2bit-approx-subchannels}
        \tildeV{i}{m}(y_1^n, u_1^{i-1} | u_i, u_{i+1}) = \max_{u_{i+2}^n \in \Bin^{n-i-1}} \W{}{m}(y_1^n | u_1^n)
\end{equation}
with $1 \leq i < n$. By definition one has
\begin{subequations}
\begin{equation}
\begin{aligned}
\label{eq:boundary-approx-channel-1}
        \tildeW{i}{m} & (y_1^n, u_1^{i-1} | u_i) = \max_{u_{i+1}^n \in \Bin^{n-i}} \W{}{m}(y_1^n | u_1^n) \\
        &= \max_{u_{i+1} \in \Bin} \max_{u_{i+2}^n \in \Bin^{n-i-1}} \W{}{m}(y_1^n | u_1^n) \\
        &= \max_{u_{i+1} \in \Bin} \tildeV{i}{m}(y_1^n, u_1^{i-1} | u_i, u_{i+1}),
\end{aligned}
\end{equation}
\begin{equation}
\label{eq:boundary-approx-channel-2}
        \tildeW{i+1}{m}(y_1^n, u_1^i | u_{i+1}) = \tildeV{i}{m}(y_1^n, u_1^{i-1} | u_i, u_{i+1})
\end{equation}
\end{subequations}
establishing the relation between the single- and double-bit virtual channels. Next, for $1 \leq i < n$ let us define LLRs for these approximate probabilities:
\begin{subequations}
\begin{align}
        \LL{i}{m}(y_1^n, u_1^{i-1}) &= \ln\frac{\max\limits_{b \in \Bin} \tildeV{i}{m}(y_1^n,u_1^{i-1}|0,b)}{\max\limits_{b \in \Bin} \tildeV{i}{m}(y_1^n,u_1^{i-1}|1,b)}, \\
        \RR{i}{m}(y_1^n, u_1^{i-1};b) &= \ln\frac{\tildeV{i}{m}(y_1^n, u_1^{i-1}|b,0)}{\tildeV{i}{m}(y_1^n, u_1^{i-1}|b,1)},\ b \in \Bin.
\end{align}
\end{subequations}
So, instead of computing $\V{i}{\lambda}(\hat x_{i,\beta}^{(\lambda)} | a,b),\ a,b \in \Bin$ for $1 \leq i < 2^{\lambda}$ and $1 \leq \beta \leq 2^{m-\lambda}$, the proposed decoder maintains three real numbers $\LL{i}{\lambda}(\hat x_{i,\beta}^{(\lambda)})$, $\RR{i}{\lambda}(\hat x_{i,\beta}^{(\lambda)};0)$ and $\RR{i}{\lambda}(\hat x_{i,\beta}^{(\lambda)};1)$. It can be seen that \eqref{eq:boundary-approx-channel-1} and \eqref{eq:boundary-approx-channel-2} imply
\begin{equation}
\label{eq:prev-approx-llr}
\begin{aligned}
        \LL{i}{m}(y_1^n,u_1^{i-1}) =& \RR{i-1}{m}(y_1^n,u_1^{i-2};u_{i-1}) \\
        =& \ln\frac{\tildeW{i}{m}(y_1^n,u_1^{i-1}|0)}{\tildeW{i}{m}(y_1^n,u_1^{i-1}|1)}
\end{aligned}
\end{equation}
what represents the LLR for the $i$-th input bit. In what follows we often omit the arguments and use the shortcuts $\LL{i}{\lambda} \triangleq \{\LL{i}{\lambda}(\hat x_{i,\beta}^{(\lambda)})\}_{\beta=1}^{2^{m-\lambda}}$ and $\RR{i}{\lambda} \triangleq \{ \RR{i}{\lambda}(\hat x_{i,\beta}^{(\lambda)};0), \RR{i}{\lambda}(\hat x_{i,\beta}^{(\lambda)};1) \}_{\beta = 1}^{2^{m-\lambda}}$.

Now let us introduce the recursive relation between $\{\tildeV{i}{\lambda + 1}\}_{i = 1}^{2^{\lambda+1}-1}$ and $\{\tildeV{j}{\lambda}\}_{j = 1}^{2^\lambda-1}$.

\begin{definition}
\label{def:approx-subchannels}
Consider a channel $V: \Bin^2 \to \mY$ and $s \in \{-1,0,1\},\ t \in \{0,1,2\}$. For $y_1^2 \in \mY^2,\ u_1^4 \in \Bin^4$, let
\begin{equation*}
        \nu(y_1^2, u_1^4) = V(y_1|u_1+u_2,u_3+u_4)V(y_2|u_2,u_4).
\end{equation*}
Then, one has
\begin{equation}
        V_s^{(t)}(y_1^2,u_1^t|u_{t+1},u_{t+2}) = \max\limits_{u_{t+3}^4} \nu(y_1^2, \tilde u_1^4),
\end{equation}
where for any $u_1^4 \in \Bin^4$ the vector $\tilde u_1^4 \in \Bin^4$ is defined such that $\tilde u_1 = u_1,\ \tilde u_4 = u_4,\ (\tilde u_2, \tilde u_3) = \tau^{(s)}(u_2,u_3)$ and
\begin{align*}
        \tau^{(-1)}(a,b) &= (a,b),\\
        \tau^{(0)}(a,b) &= (b,a),\\
        \tau^{(1)}(a,b) &= (a+b,b).
\end{align*}
\end{definition}

The values defined above approximate transition probabilities of the virtual channels defined in \cite{li2024abs} by considering the probability of the most likely input vector to be transmitted. In fact, $V_s^{(0)},V_s^{(1)},V_s^{(2)}$ for $s=-1$ approximate probabilities of $V^\oria,V^\orib,V^\oric$; for $s=0$ they approximate probabilities of $V^\swpa,V^\swpb,V^\swpc$ and for $s=1$ they approximate probabilities of $V^\adda,V^\addb,V^\addc$ (see \cite{li2024abs}). For the convenience, in what follows we often consider $V_s^{(t)}$'s as DBI channels.

Based on the transforms from Definition \ref{def:approx-subchannels}, the following lemma allows one to compute approximate transition probabilities $\tildeV{i}{m}$ from \eqref{eq:2bit-approx-subchannels}.

\begin{lemma}
\label{lm:approx-channels}
Let us fix the matrices $\QABS{1}, \QABS{2}, \dots, \QABS{m},\ m \ge 2$ according to \eqref{eq:qabs-def} with the sets $\IS{m}$ and $\IA{m}$ satisfying the conditions (\ref{item:cond-1})-(\ref{item:cond-2}) from Section \ref{subsect:abs-polar-codes}. For every $i:\ 1 \leq i < n/2$ one has
\begin{enumerate}[(i)]
        \item \label{item:swp-case} If $2i \in \IS{m}$, then for any $t \in \{0,1,2\}$
        \begin{equation*}
                \tildeV{2i+t-1}{m} = (\tildeV{i}{m-1})^{(t)}_0.
        \end{equation*}
        \item \label{item:add-case} If $2i \in \IA{m}$, then for any $t \in \{0,1,2\}$
        \begin{equation*}
                \tildeV{2i+t-1}{m} = (\tildeV{i}{m-1})^{(t)}_1.
        \end{equation*}
        \item \label{item:ori-case-1} If $2(i-1) \in \IS{m},\ 2(i+1) \in \II{m}$, then $$\tildeV{2i}{m} = (\tildeV{i}{m-1})^{(1)}_{-1}.$$
        \item \label{item:ori-case-2} If $2(i-1) \in \IS{m},\ 2(i+1) \not\in \II{m}$, then
        \begin{equation*}
                \tildeV{2i}{m} = (\tildeV{i}{m-1})^{(1)}_{-1},\
                \tildeV{2i+1}{m} = (\tildeV{i}{m-1})^{(2)}_{-1}.
        \end{equation*}
        \item \label{item:ori-case-3} If $2(i-1) \not\in \IS{m},\ 2(i+1) \in \II{m}$, then
        \begin{equation*}
                \tildeV{2i-1}{m} = (\tildeV{i}{m-1})^{(0)}_{-1},\ \tildeV{2i}{m} = (\tildeV{i}{m-1})^{(1)}_{-1}.
        \end{equation*}
        \item \label{item:ori-case} If $2(i-1) \not\in \IS{m},\ 2i,2(i+1) \not\in \II{m}$, then for any $t \in \{0,1,2\}$
        \begin{equation*}
                \tildeV{2i+t-1}{m} = (\tildeV{i}{m-1})^{(t)}_{-1}.
        \end{equation*}
\end{enumerate}
\end{lemma}

We discuss the proof of this lemma in Appendix \ref{app:proof-lm-1}.

In Lemma \ref{lm:approx-llrs}, we present the formulae for recursive calculation of approximate LLRs for virtual DBI subchannels.

\begin{lemma}
\label{lm:approx-llrs}
Consider a DBI channel $V: \Bin \to \mY$ and let
\begin{equation*}
        \label{eq:2bit-llr}
        L(y) = \ln\frac{\max\limits_{b \in \Bin}V(y|0,b)}{\max\limits_{b \in \Bin}V(y|1,b)},\ R(y;b) = \ln\frac{V(y|b,0)}{V(y|b,1)},
\end{equation*}
where $y \in \mY$, $b \in \Bin$. Next, let $y_1,y_2 \in \mY$,\ $u_1,u_2,u_3 \in \Bin$, $s \in \{-1,0,1\},\ t \in \{0,1,2\}$ and
\begin{align*}
        &L_s^{(t)}(y_1^2,u_1^t) = \ln\frac{\max\limits_{b \in \Bin} V_s^{(t)}(y_1^2,u_1^t|0,b)}{\max\limits_{b \in \Bin} V_s^{(t)}(y_1^2,u_1^t|1,b)},\\
        &R_s^{(t)}(y_1^2,u_1^t;u_{t+1}) = \ln\frac{V_s^{(t)}(y_1^2,u_1^t|u_{t+1},0)}{V_s^{(t)}(y_1^2,u_1^t|u_{t+1},1)}.
\end{align*}
        
Then, one has
{\allowdisplaybreaks
\begin{subequations}
\label{eq:llr-transform}
\begin{equation}
        L_s^{(0)}(y_1^2) = f_-(L(y_1),L(y_2)), \label{eq:llr-oria}
\end{equation}
\begin{equation}
        L_{-1}^{(1)}(y_1^2,u_1) = R_{-1}^{(0)}(y_1^2;u_1) = f_+(L(y_1),L(y_2),u_1), \label{eq:llr-orib}
\end{equation}
\begin{equation}
        \begin{aligned}
                L_s^{(1)} & (y_1^2,u_1) = R_s^{(0)}(y_1^2;u_1) = g((-1)^s R_{-1}^{(1)}(y_1^2,u_1;1), \\ & R_s^{(1)}(y_1^2,u_1;0), R_s^{(1)}(y_1^2,u_1;1)),\ s \not= -1
        \end{aligned} \label{eq:llr-b}
\end{equation}
\begin{equation}
        \begin{aligned}
                L_{-1}^{(2)}(y_1^2,u_1^2) =& R_{-1}^{(1)}(y_1^2,u_1;u_2) = \\ =& f_-(R(y_1;u_1+u_2),R(y_2;u_2)),
        \end{aligned} \label{eq:llr-oric}
\end{equation}
\begin{equation}
        \begin{aligned}
                L_s^{(2)} & (y_1^2,u_1^2) = R_s^{(1)}(y_1^2,u_1;u_2) = (-1)^{su_2} g(L_{-1}^{(1)}(y_1^2,u_1), \\ & (-1)^{u_2+1} R_{-1}^{(1)}(y_1^2,u_1;0), (-1)^{u_2+s+1} R_{-1}^{(1)}(y_1^2,u_1;1)), \\ & s \not= -1,
        \end{aligned} \label{eq:llr-c-0}
\end{equation}
\begin{equation}
        R_{-1}^{(2)}(y_1^2,u_1^2;u_3) = f_+(R(y_1;u_1+u_2), R(y_2;u_2), u_3), \label{eq:llr-oric-right}
\end{equation}
\begin{equation}
        R_0^{(2)}(y_1^2,u_1^2;u_3) = f_+(R(y_1;u_1+u_3), R(y_2;u_3), u_2), \label{eq:llr-swpc-right}
\end{equation}
\begin{equation}
        \begin{aligned}
                R_1^{(2)} & (y_1^2,u_1^2;u_3) = \\ =& f_+(R(y_1;u_1+u_2+u_3), R(y_2;u_2+u_3), u_3).
        \end{aligned} \label{eq:llr-addc-right}
\end{equation}
\end{subequations}}
with the functions $f_+$ and $f_-$ defined in Section \ref{subsect:arikan-sc-decoding} and $g(a,b,c) = a - \max\{0,b\} + \max\{0,c\}$.
\end{lemma}

The proof of this lemma is given in Appendix \ref{app:proof-lm-2}.

Similarly to probability-domain decoding, the LLR-based version of the SC (list) algorithm computes approximate LLRs for the input bits. For this, we suggest to recursively apply the formulae \eqref{eq:llr-oria}-\eqref{eq:llr-addc-right}. More precisely, consider processing of the node $(\lambda,i)$ (so that the decoder has $\LL{i}{\lambda}$ and $\RR{i}{\lambda}$ computed) and some $1 \leq \beta \leq N = 2^{m-\lambda-1}$. Let us also assign $V = \tildeV{i}{\lambda}$, $y_1 = \hat x_{i,\beta}^{(\lambda)},\ y_2 = \hat x_{i,\beta + N}^{(\lambda)},\ u_1 = \hat u_{2i-1,\beta}^{(\lambda+1)},\ u_2 = \hat u_{2i,\beta}^{(\lambda+1)},\ u_3 = \hat u_{2i+1,\beta}^{(\lambda+1)}$. Then, in terms of the notations of Lemma \ref{lm:approx-llrs}, one has $L(y_1) = \LL{i}{\lambda}(\hat x_{i,\beta}^{(\lambda)}),\ R(y_1;b) = \RR{i}{\lambda}(\hat x_{i,\beta}^{(\lambda)};b),\ L(y_2) = \LL{i}{\lambda}(\hat x_{i,\beta + N}^{(\lambda)}),\ R(y_2;b) = \RR{i}{\lambda}(\hat x_{i,\beta + N}^{(\lambda)};b),\ b \in \Bin$ are the known values. Hence, according to Lemma \ref{lm:approx-channels}, $\LL{2i-1}{\lambda+1}(\hat x_{2i-1,\beta}^{(\lambda+1)})$ can be calculated as \eqref{eq:llr-oria}, $\LL{2i}{\lambda+1}(\hat x_{2i,\beta}^{(\lambda+1)}) = \RR{2i-1}{\lambda+1}(\hat x_{2i-1,\beta}^{(\lambda+1)}; \hat u_{2i-1,\beta}^{(\lambda+1)})$ can be calculated as \eqref{eq:llr-orib} or \eqref{eq:llr-b}, $\LL{2i+1}{\lambda+1}(\hat x_{2i+1,\beta}^{(\lambda+1)}) = \RR{2i}{\lambda+1}(\hat x_{2i,\beta}^{(\lambda)}; \hat u_{2i,\beta}^{(\lambda+1)})$ can be calculated as \eqref{eq:llr-oric} or \eqref{eq:llr-c-0} and finally $\RR{2i+1}{\lambda+1}(\hat x_{2i+1,\beta}^{(\lambda+1)}; \hat u_{2i+1,\beta}^{(\lambda+1)})$ can be calculated using one of the formulae \eqref{eq:llr-oric-right}-\eqref{eq:llr-addc-right}.

In the following sections, we introduce some ideas that might significantly reduce the computational cost of the LLRs evaluation and present the formal description of the LLR-based decoding algorithm.

\subsection{Reusing intermediate values for LLR computations} \label{subsect:reusing}

In this section, we show that naive application of the formulae from Lemma \ref{lm:approx-llrs} in the decoder leads to performing redundant arithmetic operations. In those cases the decoder obtains exactly the same values twice. Instead of that, it might store common values (i.e. values that are needed for processing of different phases) and copy them instead of second-time evaluation.

Namely, \eqref{eq:prev-approx-llr} implies that instead of computing $\LL{i}{\lambda}(\hat x_{i,\beta}^{(\lambda)})$ and the $i$-th phase, one can reuse the value $\RR{i-1}{\lambda}(\hat x_{i-1,\beta}^{(\lambda)}; \hat u_{i-1,\beta}^{(\lambda)})$ with the bit $\hat u_{i-1,\beta}^{(\lambda)}$ having been decoded at the $(i-1)$-th phase.

Moreover, according to \eqref{eq:llr-b}, for $s \in \Bin$ the computation of $R_s^{(0)}(y_1^2;u_1)$ requires $R_s^{(1)}(y_1^2,u_1;0)$ and $R_s^{(1)}(y_1^2,u_1;1)$ to be known. So, if $2i \in \II{\lambda+1}$, the values $\RR{2i}{\lambda+1}(\hat x_{2i,\beta}^{(\lambda+1)};b),\ 1 \leq \beta \leq 2^{m-\lambda-1},\ b \in \Bin$ are being obtained while calculating $\RR{2i-1}{\lambda+1}(\hat x_{2i-1,\beta}^{(\lambda+1)};b),\ 1 \leq \beta \leq 2^{m-\lambda-1},\ b \in \Bin$ and might be simply reused at the $(2i)$-th phase of the layer $\lambda+1$.

The formal explanation of how these observations are injected in the decoding algorithm is given in Section \ref{subsect:llr-algorithm}.
 
\subsection{Omitting computations of unused LLRs} \label{subsect:unused-llrs}

It can be seen that the original decoding algorithm \cite{li2024abs} of ABS+ polar codes keeps unnecessary information about transition probabilities of virtual subchannels. Namely, at the boundary layer it gets the probabilities for the $i$-th and the $(i+1)$-th input bits implicitly evaluated for some $1 \leq i < n$. Here "implicitly" means that the decoder does not store the probabilities for $u_i$ and $u_{i+1}$ to be $0$ or $1$, but the values $\mathbb{P}(u_i=a,u_{i+1}=b),\ a,b\in\Bin$, which straightforwardly define the single-bit probabilities. However, if $i < n-1$, the algorithm decodes only the $i$-th bit while the $(i+1)$-th one is being processed only at the next phase. Instead of that, one can decode both bits at the $i$-th phase.

In the case of LLR-domain decoding, this approach can be seen in even simpler way. To decode the $i$-th input bit, it is sufficient to evaluate $\LL{i}{m}(y_1^n, \hat u_1^{i-1})$ and make the hard decision without maintaining $\RR{i}{m}(y_1^n, \hat u_1^{i-1};b),\ b \in \Bin$. Considering layers other than the $m$-th one, we are to show that in many cases for $0 \leq \lambda < m,\ 1 \leq i \leq 2^\lambda$ the decoder might get $\hat u_i^{(\lambda)}$ without computing the values $\RR{i}{\lambda}$.

For the convenience of the discussion, let us reformulate the SC algorithm from \cite{li2024abs}. Here we give only high-level explanation, which is necessary for understanding which LLR computations can be neglected. More formal algorithm description is given in Section \ref{subsect:llr-algorithm}. Unlike the original one, our version of the algorithm processes $m+1$ layers (numbered from zero) and $2^\lambda$ phases at each layer $\lambda$ (instead of $2^{\lambda}-1$). After the node $(\lambda,i)$ has been processed, the algorithm has $u_i^{(\lambda)}$ decoded. If $\lambda=m$, it just makes the hard decision with respect to $\LL{i}{m}$. Otherwise, if $2i \not\in \II{\lambda+1}, 2(i-1) \not\in \II{\lambda+1}$, the nodes $(\lambda+1,2i-1)$ and $(\lambda+1,2i)$ are being recursively processed. If $2(i-1) \in \II{\lambda+1}
$, the algorithm processes only the node $(\lambda+1,2i)$ and uses previously stored value of $\hat u_{2i-1}^{(\lambda+1)}$. If $2i \in \II{\lambda+1}$, the algorithm additionally processes the node $(\lambda+1,2i+1)$. If $2i \in \IS{\lambda+1}$, $\hat u_{2i}^{(\lambda+1)}$ is being stored and $\hat u_i^{(\lambda)}$ is calculated according to \eqref{eq:abs-component-vectors-3}. If $2i \in \IA{\lambda+1}$, $\hat u_{2i+1}^{(\lambda+1)}$ is being stored and $\hat u_i^{(\lambda)}$ is calculated according to \eqref{eq:abs-component-vectors-4}. As the input for processing of the node $(\lambda,i)$ the algorithm receives $\LL{i}{\lambda}, \RR{i}{\lambda}$ if $i < 2^\lambda$ and only $\LL{i}{\lambda}$ if $i=2^\lambda$. In the case $i < 2^\lambda$, the decoder uses Lemmas \ref{lm:approx-channels} and \ref{lm:approx-llrs} to calculate $\LL{i}{\lambda}$ and $\RR{i}{\lambda}$. If $i=2^\lambda$, the decoder reuses $\RR{i-1}{\lambda}$: $\LL{i}{\lambda}(\hat x_{i,\beta}) = \RR{i-1}{\lambda}(\hat x_{i,\beta}^{(\lambda)}; \hat u_{i-1,\beta}^{(\lambda)}), 1 \leq \beta \leq 2^{m-\lambda}$. The initial values in this recursive procedure are $\LL{1}{0}(y_\beta) = \ln\dfrac{W(y_\beta|0)}{W(y_\beta|1)},\ 1 \leq \beta \leq n$. Finally, since $2^{\lambda+1}$ never belongs to $\II{\lambda+1}$, for computing $\LL{2^{\lambda+1}}{\lambda+1}$ it is sufficient to know only $\LL{2^\lambda}{\lambda}$ and $\hat u_{2^\lambda-1}^{(\lambda+1)}$ (see \eqref{eq:llr-orib}). An example of the recursion tree in the reformulated SC decoder is given in Fig. \ref{fig:abs-sc-tree-2}.

Note that for all $0 \leq \lambda \leq m$ the node $(\lambda,2^\lambda)$ can be processed if only one LLR $\LL{2^\lambda}{\lambda}$ is known. But there may be some other nodes requiring only such single-bit LLRs for their processing. So, let us investigate the cases when $\RR{i}{\lambda}$ is indeed needed for decoding of $u_i^{(\lambda)}$. The key observation is that if $2i \not\in \II{\lambda+1}$, according to \eqref{eq:llr-oria} and \eqref{eq:llr-orib}, one can obtain $\LL{2i-1}{\lambda+1},\ \RR{2i-1}{\lambda+1}$ and $\LL{2i}{\lambda}$ by taking into account only the values $\LL{i}{\lambda}$ (and also $u_{2i-1}^{(\lambda)}$ which is being decoded at the $(2i-1)$-th phase of the layer $\lambda+1$). Note that in this case $\hat u_{2i+1}^{(\lambda+1)}$ is not required for decoding of $u_i^{(\lambda)}$ (see \eqref{eq:abs-component-vectors-1}). However, the "right" component vector in the decomposition of $u_{2i}^{(\lambda+1)}$ may require $\RR{2i}{\lambda+1}$ in order to be decoded. To calculate these LLRs, the decoder must have $\RR{i}{\lambda}$ computed (see \eqref{eq:llr-oric-right}-\eqref{eq:llr-addc-right}). So, for $1 \leq i < 2^\lambda$ the algorithm does not use $\RR{i}{\lambda}$ for processing of the node $(\lambda,i)$ if both conditions hold:
\begin{enumerate}
        \item \label{item:right-separated-1} $2i \not\in \II{\lambda+1}$;
        \item \label{item:right-separated-2} $\lambda=m$ or the algorithm does not use $\RR{2i}{\lambda+1}$ in order to process the node $(\lambda+1,2i)$.
\end{enumerate}

\begin{figure}
        \centering
        \begin{tikzpicture}[-{Stealth[length=2mm]}]

\node (u01) at (0,0) {$(0,1)$};

\node (u11) at (-2,-1) {$(1,1)$};
\node (u12) at (2,-1) {$(1,2)$};

\node (u21) at (-3,-2) {$(2,1)$};
\node (u22) at (-1,-2) {$(2,2)$};
\node (u23) at (1,-2) {$(2,3)$};
\node (u24) at (3,-2) {$(2,4)$};

\node (u31) at (-3.5,-3) {$(3,1)$};
\node (u32) at (-2.5,-3) {$(3,2)$};
\node (u33) at (-1.5,-3) {$(3,3)$};
\node (u34) at (-0.5,-3) {$(3,4)$};
\node (u35) at (0.5,-3) {$(3,5)$};
\node (u36) at (1.5,-3) {$(3,6)$};
\node (u37) at (2.5,-3) {$(3,7)$};
\node (u38) at (3.5,-3) {$(3,8)$};

\draw (u01) edge (u11);
\draw (u01) edge (u12);

\draw[red] (u11) edge (u21);
\draw[red] (u11) edge (u22);
\draw[red] (u11) edge (u23);
\draw (u12) edge (u24);

\draw (u21) edge (u31);
\draw (u21) edge (u32);
\draw[red] (u22) edge (u33);
\draw[red] (u22) edge (u34);
\draw[red] (u22) edge (u35);
\draw (u23) edge (u36);
\draw (u24) edge (u37);
\draw (u24) edge (u38);

\end{tikzpicture}
        \caption{The tree of the recursion in the proposed SC algorithm applied to an ABS+ polar code with $\IS{2} = \emptyset,\ \IA{2} = \{2\},\ \IS{3} = \{4\},\ \IA{3} = \emptyset$. The algorithm performs depth-first-search starting from the root node $(0,1)$; children are being processed from left to right. We use a black edge from $(\lambda,i)$ to $(\lambda+1,i')$ if $2i \not\in \II{\lambda+1}$. Otherwise the edge is red.}
        \label{fig:abs-sc-tree-2}
\end{figure}

Let us give a more precise explanation of those conditions. While processing of the node $(\lambda,i)$ the decoder often estimates not only the vector $u_i^{(\lambda)}$, but also some component vectors in the decomposition of $u_{i+1}^{(\lambda)}$. Namely, if $2i \in \II{\lambda+1}$, it is necessary to decode all vectors $u_{2i-1}^{(\lambda+1)}$, $u_{2i}^{(\lambda+1)}$ and $u_{2i+1}^{(\lambda+1)}$, although their mapping to $u_i^{(\lambda)}$ is not bijective. Then, one of the decoded vectors (i.e. $u_{2i}^{(\lambda+1)}$ when $2i \in \IS{\lambda+1}$ and $u_{2i+1}^{(\lambda+1)}$ when $2i \in \IA{\lambda+1}$) is being stored, because it is needed for processing of the node $(\lambda,i+1)$. Moreover, even if $2i \not\in \II{\lambda+1}$, some of rightmost descendants $(\lambda',i')$ of $(\lambda,i)$ may satisfy $2i' \in \II{\lambda'+1}$, so that for processing $(\lambda',i')$ one needs to have $\RR{\lambda'}{i'}$ computed. So, both conditions \ref{item:right-separated-1}) and \ref{item:right-separated-2}) hold iff all the edges on the rightmost path from $(\lambda,i)$ to a leaf node have the form $(\lambda',i') \to (\lambda'+1,2i')$. In what follows we call such nodes $(\lambda,i)$ \textit{right-separated}. Note that after the node $(\lambda,i)$ has been processed, one has the input bits $u_1,u_2,\dots,u_j$ estimated where $(m,j)$ is the rightmost leaf node in the subtree of $(\lambda,i)$. Therefore, if $(\lambda,i)$ \textit{is not} right-separated, there are some input bits which are being decoded at the current phase and whose values restrict the set of all possible values of $u_{i+1}^{(\lambda)}$. In these cases we indeed need to take into account some information about bits of $u_{i+1}^{(\lambda)}$, which is represented by $\RR{i}{\lambda}$. For instance, in the SC decoding tree from Fig. \ref{fig:abs-sc-tree-2}, right-separated nodes are $(0,1), (1,2), (2,1), (2,3), (2,4)$ and all the leaf nodes.

\begin{algorithm}[b]
        \caption{SC$(y_1,y_2,\dots,y_n)$}
        \label{alg:sc-main}
        
        \For{$\beta \in \{1,2,\dots,n\}$}{
                $\LeftLLR[0,\beta] \gets \ln\dfrac{W(y_\beta|0)}{W(y_\beta|1)}$\;
        }
        DecodeNode$(0,1)$\;
        \For{$\beta \in \{1,2,\dots,n\}$}{
                $\hat c_\beta \gets \Res[0,\beta]$\;
        }
        \Return $\hat c_1^n$\;
\end{algorithm}

\subsection{The proposed LLR-based SC decoding algorithm} \label{subsect:llr-algorithm}

Let us move to the formal definition of the proposed SC decoding algorithm. The main endpoint is presented in Algorithm \ref{alg:sc-main}, where the recursive procedure DecodeNode from Algorithm \ref{alg:sc-decode-node} is called to process the root node. The function DecodeNode$(\lambda,i)$ processes the node $(\lambda,i)$ in the way discussed above. To explore it precisely, consider the underlying data structures.

\begin{algorithm}
        \caption{DecodeNode$(\lambda,i)$}
        \label{alg:sc-decode-node}
        
        \If{$\lambda = m$}{
                $\Res[m,1] \gets 1\ \textbf{if}\ \LeftLLR[m,1] < 0\ \textbf{and}\ i \not\in \Frozen\ \textbf{else}\ 0$\; \label{line:hard-decision}
                \Return\;
        }
        $N \gets 2^{m-\lambda}$\;
        \eIf{$2(i-1) \not\in \II{\lambda+1}$}{
                \For{$\beta \in \{1,2,\dots,N/2\}$}{
                        CalculateLLRs\_LeftBranch$(\lambda,i,\beta)$\; \label{line:calc-llrs-left}
                }
                DecodeNode$(\lambda+1,2i-1)$\;
                \For{$\beta \in \{1,2,\dots,N/2\}$}{
                        $\Res[\lambda,\beta] \gets \Res[\lambda+1,\beta]$\;
                }
        }{
                \For{$\beta \in \{1,2,\dots,N/2\}$}{
                        $\Res[\lambda,\beta] \gets \Hlp[\lambda,\beta]$\; \label{line:restore}
                }
        }
        \For{$\beta \in \{1,2,\dots,N/2\}$}{
                CalculateLLRs\_MiddleBranch$(\lambda,i,\beta)$\; \label{line:calc-llrs-middle}
        }
        DecodeNode$(\lambda+1,2i)$\;
        \eIf{$2i \not\in \II{\lambda+1}$}{
                \For{$\beta \in \{1,2,\dots,N/2\}$}{
                        $\beta' \gets \beta+N/2,\ r_1 \gets \Res[\lambda,\beta],\ r_2 \gets \Res[\lambda+1,\beta]$\;
                        $\Res[\lambda,\beta] \gets r_1+r_2,\ \Res[\lambda,\beta'] \gets r_2$\; \label{line:get-result-ori}
                }
        }{
                \For{$\beta \in \{1,2,\dots,N/2\}$}{
                        $\Res[\lambda,\beta+N/2] \gets \Res[\lambda+1,\beta]$\;
                }
                \For{$\beta \in \{1,2,\dots,N/2\}$}{
                        CalculateLLRs\_RightBranch$(\lambda,i,\beta)$\; \label{line:calc-llrs-right}
                }
                DecodeNode$(\lambda+1,2i+1)$\;
                \For{$\beta \in \{1,2,\dots,N/2\}$}{
                        $\beta' \gets \beta + N/2,\ r_1 \gets \Res[\lambda,\beta],\ r_2 \gets \Res[\lambda,\beta'],\ r_3 \gets \Res[\lambda+1,\beta]$\;
                        \If{$2i \in \IS{\lambda+1}$}{
                                $\Res[\lambda,\beta] \gets r_1+r_3,\ \Res[\lambda,\beta'] \gets r_3$\; \label{line:get-result-swp}
                                $\Hlp[\lambda,\beta] \gets r_2$\; \label{line:store-swp}
                        }
                        \If{$2i \in \IA{\lambda+1}$}{
                                $\Res[\lambda,\beta] \gets r_1+r_2+r_3,\ \Res[\lambda,\beta'] \gets r_2+r_3$\; \label{line:get-result-add}
                                $\Hlp[\lambda,\beta] \gets r_3$\; \label{line:store-add}
                        }
                }
        }
\end{algorithm}

\begin{itemize}
        \item 2-dimensional array $\Res$ of decoded bits. For $0 \leq \lambda \leq m,\ 1 \leq i \leq 2^\lambda,\ 1 \leq \beta \leq 2^{m-\lambda}$ after returning from DecodeNode$(\lambda,i)$, $\Res[\lambda,\beta]$ stores the estimated bit $\hat u_{i,\beta}^{(\lambda)}$.
        \item 2-dimensional array $\Hlp$ of decoded bits that are needed for processing of a next phase. For $0 \leq \lambda < m,\ 1 \leq i \leq 2^{\lambda},\ 1 \leq \beta \leq 2^{m-\lambda-1}$ after returning from DecodeNode$(\lambda,i)$, $\Hlp[\lambda,\beta]$ stores either $\hat u_{2i,\beta}^{(\lambda+1)}$ if $2i \in \IS{\lambda+1}$ or $\hat u_{2i+1,\beta}^{(\lambda+1)}$ if $2i \in \IA{\lambda+1}$.
        \item 2-dimensional array $\LeftLLR$ of real numbers representing $\LL{i}{\lambda}$. For $0 \leq \lambda \leq m,\ 1 \leq i \leq 2^\lambda,\ 1 \leq \beta \leq 2^{m-\lambda}$ before calling DecodeNode$(\lambda,i)$, $\LeftLLR[\lambda,\beta]$ stores $\LL{i}{\lambda}(\hat x_{i,\beta}^{(\lambda)})$.
        \item 2-dimensional array $\RightLLR$ of pairs of real numbers representing $\RR{i}{\lambda}$. Consider $0 \leq \lambda \leq m,\ 1 \leq i \leq 2^\lambda,\ 1 \leq \beta \leq 2^{m-\lambda},\ b \in \Bin$ such that the node $(\lambda,i)$ is not right-separated. After calling DecodeNode$(\lambda,i)$,  $\RightLLR[\lambda,\beta][b]$ stores $\RR{i}{\lambda}(\hat x_{i,\beta}^{(\lambda)};b)$.
        \item 2-dimensional array $\Mem$ of quadruples of real numbers representing the intermediate values that can be reused while LLR computations (see Section \ref{subsect:reusing}). Consider $0 \leq \lambda < m,\ 1 \leq i \leq 2^\lambda,\ 1 \leq \beta \leq 2^{m-\lambda-1}$ such that $2i \in \II{\lambda+1}$. If the node $(\lambda+1,2i-1)$ is not right-separated, after $\RR{2i-1}{\lambda+1}(\hat x_{2i-1,\beta}^{(\lambda+1)};a),\ a \in \Bin$ have been computed, $\Mem[\lambda,\beta][a,b],\ b \in \Bin$ stores $\RR{2i}{\lambda+1}(\hat x_{2i,\beta}^{(\lambda+1)};b)$ with the assumption $\hat u_{2i,\beta}^{(\lambda+1)} = a$. If the node $(\lambda+1,2i-1)$ is right-separated but $(\lambda+1,2i)$ is not, after $\LL{2i}{\lambda+1}(\hat x_{2i,\beta})$ has been computed, $\Mem[\lambda,\beta][\hat u_{2i,\beta}^{(\lambda+1)},b],\ b \in \Bin$ stores $\RR{2i}{\lambda+1}(\hat x_{2i,\beta}^{(\lambda+1)};b)$.
\end{itemize}

\begin{algorithm}[t]
        \caption{CalculateLLRs\_LeftBranch$(\lambda,i,\beta)$}
        \label{alg:calc-llrs-left}
        
        $\beta' \gets \beta + 2^{m-\lambda-1}$\;
        \tcp{Calculate $\LL{2i-1}{\lambda+1}(\hat x_{2i-1,\beta}^{(\lambda+1)})$}
        \eIf{$i>1$ \textbf{and} $(\lambda+1,2i-2)$ is not right-separated}{
                $r_0 \gets \Res[\lambda+1,\beta]$ \tcp{$r_0 = \hat u_{2i-2,\beta}^{(\lambda+1)}$} \label{line:old-bit-1}
                $\LeftLLR[\lambda+1,\beta] \gets \RightLLR[\lambda+1,\beta][r_0]$\; \label{line:reuse-1}
        }{
                $\LeftLLR[\lambda+1,\beta] \gets f_-(\LeftLLR[\lambda,\beta], \LeftLLR[\lambda,\beta'])$\;
        }
        \If{$(\lambda+1,2i-1)$ is not right-separated}{
                \tcp{Calculate $\RR{2i-1}{\lambda+1}(\hat x_{2i-1,\beta}^{(\lambda+1)};b),\ b \in \Bin$}
                \For{$b \in \Bin$}{
                        \eIf{$2i \not\in \II{\lambda+1}$}{
                                $\RightLLR[\lambda+1,\beta][b] \gets f_+(\LeftLLR[\lambda,\beta], \LeftLLR[\lambda,\beta'], b)$\;
                        }{
                                $R_0 \gets f_-(\RightLLR[\lambda,\beta][b], \RightLLR[\lambda,\beta'][0])$\;
                                $R_1 \gets f_-(\RightLLR[\lambda,\beta][b+1], \RightLLR[\lambda,\beta'][1])$\;
                                $L \gets f_+(\LeftLLR[\lambda,\beta], \LeftLLR[\lambda,\beta'], b)$\;
                                \If{$2i \in \IS{\lambda+1}$}{
                                        $\Mem[\lambda,\beta][b,0] \gets L + \min\{0,R_0\} - \min\{0,R_1\}$ \tcp{\eqref{eq:llr-c-0}, $ s=0,\ u_2=0$} \label{line:assign-mem-swp-begin}
                                        $\Mem[\lambda,\beta][b,1] \gets L - \max\{0,R_0\} + \max\{0,R_1\}$ \tcp{\eqref{eq:llr-c-0}, $s=0,\  u_2=1$} \label{line:assign-mem-swp-end}
                                        $\RightLLR[\lambda+1,\beta][b] \gets R_1 + \max\{0,\Mem[\lambda,\beta][b,0]\}$ $- \max\{0,\Mem[\lambda,\beta][b,1]\}$ \tcp{\eqref{eq:llr-b}, $s=0$}
                                }
                                \If{$2i \in \IA{\lambda+1}$}{
                                        $\Mem[\lambda,\beta][b,0] \gets L + \min\{0,R_0\} + \min\{0,R_1\}$ \tcp{\eqref{eq:llr-c-0}, $s=1,\ u_2=0$}
\label{line:assign-mem-add-begin}
                                        $\Mem[\lambda,\beta][b,1] \gets -L + \max\{0,R_0\} + \max\{0,R_1\}$ \tcp{\eqref{eq:llr-c-0}, $s=1,\ u_2=1$} \label{line:assign-mem-add-end}
                                        $\RightLLR[\lambda+1,\beta][b] \gets -R_1 + \max\{0,\Mem[\lambda,\beta][b,0]\} + \min\{0,$ $\Mem[\lambda,\beta][b,1]\}$ \tcp{\eqref{eq:llr-b}, $s=1$}
                                }
                        }
                }
        }
\end{algorithm}

Let us move on to the definition of DecodeNode$(\lambda,i)$. In case $\lambda=m$ it is sufficient to make the hard decision with respect to the sign of LLR (see line \ref{line:hard-decision}).

If $\lambda < m$, the nodes of the layer $\lambda+1$ are being processed. In lines \ref{line:calc-llrs-left}, \ref{line:calc-llrs-middle} and \ref{line:calc-llrs-right} of the Algorithm \ref{alg:sc-decode-node} we compute the approximate LLRs for the virtual channels $\V{2i-1}{\lambda+1},\ \V{2i}{\lambda+1}$ and $\V{2i+1}{\lambda+1}$ respectively. Then, after the necessary nodes of the next layer have been processed, the resulting vector $\hat u_i^{(\lambda)}$ is calculated. Namely, if $2i \not\in \II{\lambda+1}$, in line \ref{line:get-result-ori} one of the formulae \eqref{eq:abs-component-vectors-1}-\eqref{eq:abs-component-vectors-2} is used; if $2i \in \IS{\lambda+1}$, in line \ref{line:get-result-swp} the formula \eqref{eq:abs-component-vectors-3} is used, and finally if $2i \in \IA{\lambda+1}$, we use the formula \eqref{eq:abs-component-vectors-4} in line \ref{line:get-result-add}. Moreover, if $2i \in \II{\lambda+1}$, the array $\Hlp$ is being updated (see lines \ref{line:store-swp} and \ref{line:store-add}). The stored value will be read in line \ref{line:restore} while execution DecodeNode$(\lambda,i+1)$.

\begin{algorithm}[t]
        \caption{CalculateLLRs\_MiddleBranch$(\lambda,i,\beta)$}
        \label{alg:calc-llrs-middle}
        
        $\beta' \gets \beta + 2^{m-\lambda-1},\ r_1 \gets \Res[\lambda,\beta]$ \tcp{$r_1 = \hat u_{2i-1,\beta}^{(\lambda+1)}$} \label{line:old-bit-2}
        \If{$(\lambda+1,2i-1)$ is right-separated \textbf{and} $2i \in \II{\lambda+1}$}{
                \tcp{Calculate $\RR{2i}{\lambda+1}(\hat x_{2i,\beta}^{(\lambda+1)};b),\ b \in \Bin$}
                $L \gets f_+(\LeftLLR[\lambda,\beta], \LeftLLR[\lambda,\beta'], r_1)$\;
                $R_0 \gets f_-(\RightLLR[\lambda,\beta][r_1], \RightLLR[\lambda,\beta'][0])$\;
                $R_1 \gets f_-(\RightLLR[\lambda,\beta][r_1+1], \RightLLR[\lambda,\beta'][1])$\;
                \If{$2i \in \IS{\lambda+1}$}{
                        $\Mem[\lambda,\beta][r_1,0] \gets L + \min\{0,R_0\} - \min\{0,R_1\}$ \tcp{\eqref{eq:llr-c-0}, $s=0,\ u_2=0$} \label{line:assign-mem-swp-1}
                        $\Mem[\lambda,\beta][r_1,1] \gets L - \max\{0,R_0\} + \max\{0,R_1\}$ \tcp{\eqref{eq:llr-c-0}, $s=0,\ u_2=1$} \label{line:assign-mem-swp-2}
                }
                \If{$2i \in \IA{\lambda+1}$}{
                        $\Mem[\lambda,\beta][r_1,0] \gets L + \min\{0,R_0\} + \max\{0,R_1\}$ \tcp{\eqref{eq:llr-c-0}, $s=1,\ u_2=0$} \label{line:assign-mem-add-1}
                        $\Mem[\lambda,\beta][r_1,1] \gets -L + \max\{0,R_0\} + \min\{0,R_1\}$ \tcp{\eqref{eq:llr-c-0}, $s=1,\ u_2=1$} \label{line:assign-mem-add-2}
                }
        }
        \tcp{Calculate $\LL{2i}{\lambda+1}(\hat x_{2i,\beta}^{(\lambda+1)})$}
        \eIf{$(\lambda+1,2i-1)$ is not right-separated}{
                $\LeftLLR[\lambda+1,\beta] \gets \RightLLR[\lambda+1,\beta][r_1]$\; \label{line:reuse-2}
        }{
                \If{$2i \not\in \II{\lambda+1}$}{
                        $\LeftLLR[\lambda+1,\beta] \gets f_+(\LeftLLR[\lambda,\beta], \LeftLLR[\lambda,\beta'], r_1)$\;
                }
                \If{$2i \in \IS{\lambda+1}$}{
                        $\LeftLLR[\lambda+1,\beta] \gets R_1 + \max\{0,\Mem[\lambda,\beta][r_1,0]\} - \max\{0,\Mem[\lambda,\beta][r_1,1]\}$ \tcp{\eqref{eq:llr-b}, $s=0$}
                }
                \If{$2i \in \IA{\lambda+1}$}{
                        $\LeftLLR[\lambda+1,\beta] \gets -R_1 + \max\{0,\Mem[\lambda,\beta][r_1,0]\} + \min\{0,\Mem[\lambda,\beta][r_1,1]\}$ \tcp{\eqref{eq:llr-b}, $s=1$}
                }
        }
        \If{$(\lambda+1,2i)$ is not right-separated}{
                \tcp{Calculate $\RR{2i}{\lambda+1}$}
                \eIf{$2i \not\in \II{\lambda+1}$}{
                        \tcp{see \eqref{eq:llr-oric}}
                        $\RightLLR[\lambda+1,\beta][0] \gets f_-(\RightLLR[\lambda,\beta][r_1], \RightLLR[\lambda,\beta'][0])$\;
                        $\RightLLR[\lambda+1,\beta][1] \gets f_-(\RightLLR[\lambda,\beta][r_1+1], \RightLLR[\lambda,\beta'][1])$\;
                }{
                        \tcp{$\Mem[\lambda,\beta][r_1,b],\ b\in\Bin$ are assigned in lines \text{\normalfont{\ref{line:assign-mem-swp-1}-\ref{line:assign-mem-swp-2}}} and \text{\normalfont{\ref{line:assign-mem-add-1}-\ref{line:assign-mem-add-2}}} of Algorithm \text{\normalfont{\ref{alg:calc-llrs-middle}}} or in lines \text{\normalfont{\ref{line:assign-mem-swp-begin}-\ref{line:assign-mem-swp-end}}} and \text{\normalfont{\ref{line:assign-mem-add-begin}-\ref{line:assign-mem-add-end}}} of Algorithm \text{\normalfont{\ref{alg:calc-llrs-left}}}}
                        $\RightLLR[\lambda+1,\beta][0] \gets \Mem[\lambda,\beta][r_1,0]$\;
                        $\RightLLR[\lambda+1,\beta][1] \gets \Mem[\lambda,\beta][r_1,1]$\;
                }
        }
\end{algorithm}

\begin{algorithm}
        \caption{CalculateLLRs\_RightBranch$(\lambda,i,\beta)$}
        \label{alg:calc-llrs-right}
        
        $\beta' \gets \beta + 2^{m-\lambda-1},\ r_1 \gets \Res[\lambda,\beta],\ r_2 \gets \Res[\lambda,\beta']$ \tcp{$r_1 = \hat u_{2i-1,\beta}^{(\lambda+1)},\ r_2 = \hat u_{2i,\beta}^{(\lambda+1)}$}
        $\LeftLLR[\lambda+1,\beta] \gets \Mem[\lambda,\beta][r_1,r_2]$\; \label{line:reuse-3}
        \If{$(\lambda+1,2i+1)$ is not right-separated}{
                \tcp{Calculate $\RR{2i+1}{\lambda+1}(\hat x_{2i+1,\beta}^{(\lambda+1)};b),\ b \in \Bin$}
                \For{$b \in \Bin$}{
                        \If{$2i \in \IS{\lambda+1}$}{
                                $\RightLLR[\lambda+1,\beta][b] \gets$ $f_+(\RightLLR[\lambda,\beta][r_1+b],$ $ \RightLLR[\lambda,\beta'][b], r_2)$ \tcp{\eqref{eq:llr-swpc-right}}
                        }
                        \If{$2i \in \IA{\lambda+1}$}{
                                $\RightLLR[\lambda+1,\beta][b] \gets f_+(\RightLLR[\lambda,\beta][r_1+r_2+b],$ $\RightLLR[\lambda,\beta'][r_2+b],b)$ \tcp{\eqref{eq:llr-addc-right}}
                        }
                }
        }
\end{algorithm}

In essence, the differences between our function DecodeNode$(\lambda,i)$ and the recursive procedure for decoding of ABS+ polar codes from \cite{li2024abs} are very minor. Firstly, our function accepts pairs $(\lambda,i)$ such that $0 \leq \lambda \leq m,\ 1 \leq i \leq 2^\lambda$ (instead of $1 \leq \lambda \leq m,\ 1 \leq i \leq 2^\lambda-1$). Secondly, we do not decode $u_{i+1}^{(\lambda)}$ while processing the node $(\lambda,i=2^\lambda-1)$. Instead, there are the additional node $(\lambda,2^\lambda)$ in the recursion tree. And finally, the underlying arithmetic operations performed in the LLR-domain.

Now consider the Algorithms \ref{alg:calc-llrs-left}, \ref{alg:calc-llrs-middle} and \ref{alg:calc-llrs-right} which perform evaluation of the approximate LLRs. Receiving the parent node $(\lambda,i)$ and the index $1 \leq \beta \leq 2^{m-\lambda-1}$, each algorithm assigns $\LeftLLR[\lambda+1,\beta]$ for further processing either the node $(\lambda+1,2i-1)$ (Algorithm \ref{alg:calc-llrs-left}) or $(\lambda+1,2i)$ (Algorithm \ref{alg:calc-llrs-middle}) or $(\lambda+1,2i+1)$ (Algorithm \ref{alg:calc-llrs-right}). Moreover, $\RightLLR[\lambda+1,\beta][0]$ and $\RightLLR[\lambda + 1,\beta][1]$ are being assigned if and only if the "current" node (i.e. $(\lambda+1,2i-1)$, $(\lambda+1,2i)$ and $(\lambda+1,2i+1)$ for Algorithms \ref{alg:calc-llrs-left}, \ref{alg:calc-llrs-middle} and \ref{alg:calc-llrs-right} respectively) is not right-separated. Therefore, if the "preceding" node (i.e. $(\lambda+1,2i-2)$, $(\lambda+1,2i-1)$ and $(\lambda+1,2i)$ for Algorithms \ref{alg:calc-llrs-left}, \ref{alg:calc-llrs-middle} and \ref{alg:calc-llrs-right} respectively) is not right-separated, at the beginning of the function, $\RightLLR[\lambda+1,\beta]$ already stores the desired value for $\LeftLLR[\lambda+1,\beta]$. So, it is sufficient to copy the entry of $\RightLLR[\lambda+1,\beta]$ corresponding to the last decoded bit (see line \ref{line:reuse-1} in Algorithm \ref{alg:calc-llrs-left} and line \ref{line:reuse-2} in Algorithm \ref{alg:calc-llrs-middle}). Besides, since Algorithm \ref{alg:calc-llrs-right} is performed only when $2i \in \II{\lambda+1}$, the desired value of $\LeftLLR[\lambda+1,\beta]$ is always stored in $\Mem[\lambda,\beta]$ and can be copied (see line \ref{line:reuse-3}). Otherwise, $\LeftLLR[\lambda+1,\beta]$ is computed using one of the formulae \eqref{eq:llr-oria}-\eqref{eq:llr-c-0}.

\subsection{SCL decoding in the LLR domain} \label{subsect:llr-scl}

The SCL decoder maintains the arrays $\Res, \Hlp, \LeftLLR, \RightLLR, \Mem$ for every active path using the data structures from \cite{tal2015list} or \cite{li2023adjacent}. At the $i$-th phase of the boundary layer the path metrics are being updated as \eqref{eq:path-metrics} with $\alpha = \LL{i}{m}(y_1^n, \hat u_1^{i-1})$.

The non-trivial detail is that in line \ref{line:old-bit-1} of Algorithm \ref{alg:calc-llrs-left} the entry $\Res[\lambda+1,\beta]$ may refer to another path. To address this issue, one can store an additional 2-dimensional array $\RightLLR'$ for every path, so that the decoder saves $\RR{i}{\lambda}(\hat x_{i,\beta}^{(\lambda)}; \hat u_{i,\beta}^{(\lambda)}),\ 1 \leq \beta \leq 2^{m-\lambda}$ in $\RightLLR'[\lambda,\beta]$ right after the vector $\hat u_i^{(\lambda)}$ has been obtained. And instead of line \ref{line:reuse-1} in Algorithm \ref{alg:calc-llrs-left} the decoder reads $\RightLLR'[\lambda+1,\beta]$. In total, each of $L$ arrays $\RightLLR'$ consists of $2^0 + 2^1 + \dots + 2^{m-1} = n-1$ elements. Hence, it is needed to additionally store only $L(n-1)$ real numbers.

The proposed decoder requires $O(Ln)$ memory since it maintains $O(n)$ bits and real numbers for each of $L$ paths. The time complexity is $O(Ln\log n)$.

\begin{proposition}
        The presented SCL decoder of a $k$-dimensional ABS+ polar code with $L=2^k$ performs ML decoding.
\begin{proof}

It is sufficient to show that when all $n$ input bits are decoded and the path $u_1^n$ has lower metric than $v_1^n$, the codeword $u_1^n\GABS{m}$ is more likely than $v_1^n\GABS{m}$. If this holds, the most likely codeword $u_1^n \GABS{m}$ is returned. In fact, since $L=2^k$, the path $u_1^n$ must be present in the list.

Let $0 \leq i \leq n,\ y_1^n \in \mY^n,\ u_1^i \in \Bin^i$ and
\begin{equation*}
        \MM{i}{m}(y_1^n,u_1^i) = \ln\max_{u_{i+1}^n \in \Bin^{n-i}}\W{}{m}(y_1^n|u_1^n).
\end{equation*}
Hence, $\LL{i}{m}(y_1^n,u_1^{i-1}) = \MM{i}{m}(y_1^n,u_1^{i-1}.0) - \MM{i}{m}(y_1^n,u_1^{i-1}.1)$. For some path $u_1^n$ consider $\hat u_i = \argmax\limits_{b\in\Bin} \tildeW{i}{m}(y_1^n,u_1^{i-1}|b)$. Then by definition $\MM{i}{m}(y_1^n,u_1^{i-1}.\hat u_i) = \MM{i-1}{m}(y_1^n,u_1^{i-1})$. We claim that after $i$ input bits have been decoded, the metric of the path prefix $u_1^i$ is given by
\begin{equation}
\label{eq:exact-metric}
        \PM{i} = \MM{0}{m}(y_1^n) - \MM{i}{m}(y_1^n, u_1^i).
\end{equation}
Indeed, for $i = 0$ \eqref{eq:exact-metric} equals zero and for any $1 \leq i \leq n$ one can assume that the PM at the $(i-1)$-th phase equals
\begin{equation}
\label{eq:prev-metric}
        \PM{i-1} = \MM{0}{m}(y_1^n) - \MM{i-1}{m}(y_1^n, u_1^{i-1})
\end{equation}
and consider two cases.
\begin{enumerate}
        \item If $u_i = \hat u_i$, then $\MM{i}{m}(y_1^n,u_1^i) = \MM{i-1}{m}(y_1^n,u_1^{i-1})$. On the other hand, the PM at the $i$-th phase equals the PM at the $(i-1)$-th phase which equals \eqref{eq:exact-metric}.
        \item If $u_i \not= \hat u_i$, then the PM is penalized by adding
        \begin{align*}
        |\LL{i}{m}&(y_1^n,u_1^{i-1})| = \MM{i}{m}(y_1^n,u_1^{i-1}.\hat u_i) - \MM{i}{m}(y_1^n,u_1^i),
        \end{align*}
which equals $\PM{i} - \PM{i-1}$ as defined in \eqref{eq:exact-metric} and \eqref{eq:prev-metric}.
\end{enumerate}
Finally, the PM of $u_1^n$ is $\MM{0}{m}(y_1^n) - \MM{n}{m}(y_1^n,u_1^n)$. This value decreases as far as $\W{}{m}(y_1^n|u_1^n)$ increases.
\end{proof}
\end{proposition}
\begin{remark}
        This proof can be applied to a min-sum list decoder of binary polar codes with arbitrary kernels.
\end{remark}

\section{Numeric Results} \label{sect:numeric-results}

In this section we present some simulation results in the case of AWGN channel and BPSK modulation.

\begin{figure}[t!]
        \centering
        \includegraphics[width=0.5\textwidth]{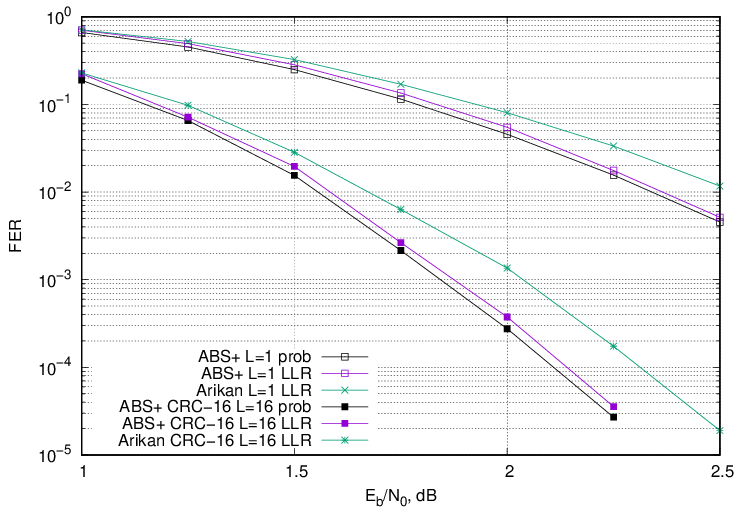}
        \caption{Performance of $(1024,512)$ polar codes.}
        \label{fig:plot-fer}
\end{figure}

\begin{figure}[t]
    \centering
    \subfloat[Performance and list size.]{
        \includegraphics[width=0.225\textwidth]{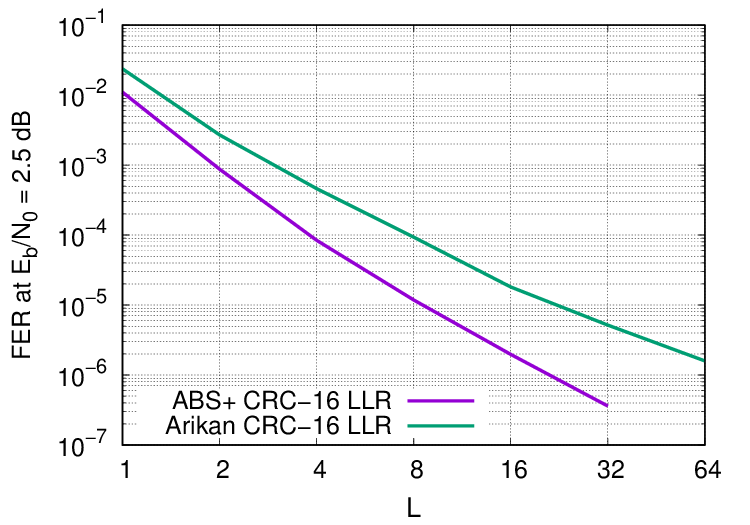}
        \label{fig:plot-list-size}
    }
    ~
    \subfloat[Performance and complexity.]{
        \includegraphics[width=0.225\textwidth]{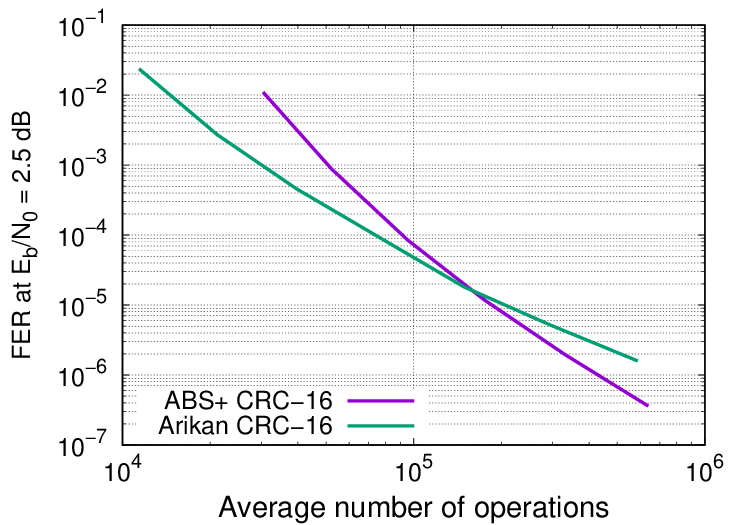}
        \label{fig:plot-complexity}
    }
    \caption{Comparison of $(1024,512)$ polar codes under SCL decoding with varied list size.}
    \label{fig:plot-performance}
\end{figure}

In Fig. \ref{fig:plot-fer} $(1024,512)$ Arikan and ABS+ polar codes are compared under SC and CRC-aided SCL decoding. In both cases the proposed LLR-based decoder of ABS+ polar codes outperforms that of the Arikan polar codes constructed using Gaussian approximation \cite{trifonov2012efficient}. The frame error rate (FER) gain is up to $0.2$ dB and increases with growing of $E_b/N_0$. Furthermore, the proposed decoder is compared against the similar probability-domain SCL ABS+ decoder. We observe a small performance loss that does not exceed $0.05$ dB.

Fig. \ref{fig:plot-list-size} represents the dependency between the list size $L$ and FER performance at the fixed $E_b/N_0 = 2.5$ dB. It can be noticed that the SCL decoder of ABS+ polar code requires much smaller list size than that of Arikan polar code in order to get the same performance. Finally, in Fig. \ref{fig:plot-complexity} we investigate the dependency between FER and the number of additions and comparisons of LLRs in the decoder. For $L \ge 12$ the ABS+ SCL decoder demonstrates better FER that the standard SCL, which uses the same number of arithmetic operations.

\section{Conclusions} \label{sect:conclusions}

In this paper, a low-complexity LLR-based version of the SCL decoder of ABS+ polar codes is discovered. This makes ABS+ polar codes applicable for a practical usage.

We introduced the approximation of LLRs for virtual subchannels stemming from the ABS+ encoding process. These LLRs, used in the proposed decoder, correspond to most likely continuations of a decoded input vector prefix. Therefore, our decoder provably achieves ML for a sufficiently large list size. Meanwhile, for a moderate list size, the loss to the similar probability-domain SCL decoder is negligibly small.

The presented min-sum approximation can be used to generalize other polar decoding approaches (e.g. sequential decoding \cite{miloslavskaya2014sequential}) to ABS+ polar codes. Moreover, one can apply the techniques of fast SC and SCL \cite{alamdaryazdi2011simplified}, \cite{hashemi2017fast} decoding to the proposed decoder.

\section{Acknowledgement}

We thank the authors of \cite{li2023adjacent} and \cite{li2024abs} for providing the software implementation of the algorithms for constructing and decoding ABS+ polar codes.
\appendix

\subsection{Proof of Lemma \ref{lm:approx-channels}} \label{app:proof-lm-1} 

Let us introduce some additional notations. We denote the $l \times l$ identity matrix as $I_l$. For random variables $\xi$ and $\eta$ we use $\mathbb{P}_{\xi | \eta}(x|y)$ to denote the conditional probability density of the distribution $\xi$ in the point $x$ with the condition $\eta = y$.

As in \cite[Proof of Lemma 1]{li2023adjacent}, consider a random vector $U_1^n$, where $U_j$ is the Bernoulli-distributed value with $\mathbb{P}_{U_j}(0) = \mathbb{P}_{U_j}(1) = 1/2$ and all $U_j$'s are independent. Next, let $X_1^n = U_1^n\GABS{m}$ and $Y_j$ is the random value corresponding to the output of $W$ if $X_j$ is given as the input. It is obvious that $W(y|x) = \mathbb{P}_{Y_j | X_j}(y|x),\ y \in \mY,\ x \in \Bin$ and therefore \begin{equation}
\label{eq:channel-prob}
        \W{}{m}(y_1^n | u_1^n) = \mathbb{P}_{Y_1^n | U_1^n}(y_1^n | u_1^n)
\end{equation}
for every $y_1^n \in \mY^n,\ u_1^n \in \Bin^n$.

Let $\tilde U_1^n = U_1^n\QABS{m}(I_{n/2} \otimes F)$. Saying informally, $\tilde U_1^n$ represents the result of passing $U_1^n$ through the leftmost layer of the encoding circuit in terms of Fig. \ref{fig:abs-factor-graph}. Since $\GABS{m} = \QABS{m}(\GABS{m-1} \otimes F) = \QABS{m}(I_{n/2} \otimes F)(\GABS{m-1} \otimes I_2)$, one has $X_1^n = \tilde U_1^n(\GABS{m-1} \otimes I_2)$ and hence $X_{1,o}^n = \tilde U_{1,o}^n\GABS{m-1}$ and $X_{1,e}^n = \tilde U_{1,e}^n\GABS{m-1}$. Therefore, the random vectors $(Y_{1,o}^n . \tilde U_{1,o})$ and $(Y_{1,e}^n . \tilde U_{1,e})$ are independent. This implies
{\allowdisplaybreaks
\begin{align*}
        \begin{split}
                \tildeV{2i-1}{m}&(y_1^n, u_1^{2i-2} | u_{2i-1}, u_{2i}) = \\ =& \max_{u_{2i+1}^n} \W{}{m}(y_1^n | u_1^n) = \max_{u_{2i+1}^n} \mathbb{P}_{Y_1^n | U_1^n}(y_1^n | u_1^n)
        \end{split} \\
        \stackrel{(a)}{=}& \max_{u_{2i+1}^n} \mathbb{P}_{Y_{1,o}^n | \tilde U_{1,o}^n}(y_{1,o}^n | \tilde u_{1,o}^n) \mathbb{P}_{Y_{1,e}^n | \tilde U_{1,e}^n}(y_{1,e}^n | \tilde u_{1,e}^n) \\
        \stackrel{(b)}{=}& \max_{u_{2i+1}^n} \W{}{m-1}(y_{1,o}^n | \tilde u_{1,o}^n) \W{}{m-1}(y_{1,e}^n | \tilde u_{1,e}^n) \\
        \stackrel{(c)}{=}& \max_{\tilde u_{2i+1}^n} \W{}{m-1}(y_{1,o}^n | \tilde u_{1,o}^n) \W{}{m-1}(y_{1,e}^n | \tilde u_{1,e}^n) \\
        =& \max_{\substack{\tilde u_{2i+1} \\ \tilde u_{2i+2}}} \max_{\tilde u_{2i+3}^n} \W{}{m-1}(y_{1,o}^n | \tilde u_{1,o}^n) \W{}{m-1}(y_{1,e}^n | \tilde u_{1,e}^n) \\
        \stackrel{(d)}{=}& \begin{aligned}[t]
                \max_{\substack{\tilde u_{2i+1} \\ \tilde u_{2i+2}}} & \left(\max_{\tilde u_{2i+3}^n} \W{}{m-1}(y_{1,o}^n | \tilde u_{1,o}^n)\right) \cdot \\ \cdot & \left( \max_{\tilde u_{2i+3}^n} \W{}{m-1}(y_{1,e}^n | \tilde u_{1,e}^n) \right)
        \end{aligned} \\
        =& \begin{aligned}[t]
                \max_{\substack{\tilde u_{2i+1} \\ \tilde u_{2i+2}}} & \tildeV{i}{m-1}(y_{1,o}^n, \tilde u_{1,o}^{2i-2} | \tilde u_{2i-1}, \tilde u_{2i+1}) \cdot \\ \cdot & \tildeV{i}{m-1}(y_{1,e}^n, \tilde u_{1,e}^{2i-2} | \tilde u_{2i}, \tilde u_{2i+2})
        \end{aligned} \\
        \stackrel{(e)}{=}& \begin{aligned}[t]
                \max_{\substack{u_{2i+1} \\ u_{2i+2}}} & 
                \tildeV{i}{m-1}(y_{1,o}^n, \tilde u_{1,o}^{2i-2} | u_{2i-1} + u_{2i}, u_{2i+1} + u_{2i+2}) \cdot \\ \cdot & \tildeV{i}{m-1}(y_{1,e}^n, \tilde u_{1,e}^{2i-2} | u_{2i}, u_{2i+2})
        \end{aligned} \\
        =& (\tildeV{i}{m-1})^{(0)}_{-1}(y_1^n, \tilde u_1^{2i-2} | u_{2i-1}, u_{2i}).
\end{align*}}

Here for $u_1^n \in \Bin^n$ we define $\tilde u_1^n = u_1^n\QABS{m}(I_{n/2} \otimes F)$. In particular, since $2(i-1) \not\in \IS{m},\ 2i,2(i+1) \not\in \II{m}$, we have
\begin{equation}
\label{eq:msg-pretransform}
\begin{aligned}
        \tilde u_{2i-1} &= u_{2i-1} + u_{2i},\ & \tilde u_{2i} &= u_{2i},\\
        \tilde u_{2i+1} &= u_{2i+1} + u_{2i+2},\ & \tilde u_{2i+2} &= u_{2i+2}.
\end{aligned}
\end{equation}
The equality $(a)$ comes from the fact that $(Y_{1,o}^n . \tilde U_{1,o})$ and $(Y_{1,e}^n . \tilde U_{1,e})$ are independent random variables and $(b)$ is the application of \eqref{eq:channel-prob}. To proof $(c)$, one needs to recall that since $2i \not\in \IS{m}$ each $u_{2i+1}^n$ is mapped bijectively to $\tilde u_{2i+1}^n$, so we can iterate over all possible vectors $\tilde u_{2i+1}^n$. The equality $(d)$ takes place since the functions $\tilde u_{2i+3}^n \mapsto \W{}{m-1}(y_{1,o}^n|\tilde u_{1,o}^n)$ and $\tilde u_{2i+3}^n \mapsto \W{}{m-1}(y_{1,e}^n|\tilde u_{1,e}^n)$ (with some fixed $y_1^n$, $u_1^{2i}$, $\tilde u_{2i+1}$ and $\tilde u_{2i+2}$) use disjoint subvectors of the input vector. The equality $(e)$ is obtained by simple relabeling $\tilde u_{2i+1} \mapsto u_{2i+1} + u_{2i+2},\ \tilde u_{2i+2} \mapsto u_{2i+2}$. As a final remark, we notice that there is the one-to-one mapping from $u_1^{2i-2}$ to $\tilde u_1^{2i-2}$ and hence the channels $\tildeV{2i-1}{m}$ and $(\tildeV{i}{m-1})^{(0)}_{-1}$ are identical.

Omitting some obvious steps, we show that $\VABS{2i}{m} = (\V{i}{m-1})^{(1)}_{-1}$ in the same way using only the fact that $u_{2i+2}^n$ is bijectively mapped to $\tilde u_{2i+2}^n$:
{\allowdisplaybreaks
\begin{align*}
        \tildeV{2i}{m}&(y_1^n, u_1^{2i-1} | u_{2i}, u_{2i+1}) = \\
        =& \max_{\tilde u_{2i+2}^n} \W{}{m-1}(y_{1,o}^n | \tilde u_{1,o}^n) \W{}{m-1}(y_{1,e}^n | \tilde u_{1,e}^n) \\
        =& \max_{\tilde u_{2i+2}} \max_{\tilde u_{2i+3}^n} \W{}{m-1}(y_{1,o}^n | \tilde u_{1,o}^n) \W{}{m-1}(y_{1,e}^n | \tilde u_{1,e}^n) \\
        =& \begin{aligned}[t]
                \max_{\tilde u_{2i+2}} & \tildeV{i}{m-1}(y_{1,o}^n, \tilde u_{1,o}^{2i-2} | \tilde u_{2i-1}, \tilde u_{2i+1}) \cdot \\ \cdot & \tildeV{i}{m-1}(y_{1,e}^n, \tilde u_{1,e}^{2i-2} | \tilde u_{2i}, \tilde u_{2i+2})
        \end{aligned} \\
        =& \begin{aligned}[t]
                \max_{u_{2i+2}} & 
                \tildeV{i}{m-1}(y_{1,o}^n, \tilde u_{1,o}^{2i-2} | u_{2i-1} + u_{2i}, u_{2i+1} + u_{2i+2}) \cdot \\
        \cdot & \tildeV{i}{m-1}(y_{1,e}^n, \tilde u_{1,e}^{2i-2} | u_{2i}, u_{2i+2})
        \end{aligned} \\
        =& (\tildeV{i}{m-1})^{(1)}_{-1}(y_1^n, \tilde u_1^{2i-2}.u_{2i-1} | u_{2i}, u_{2i+1}).
\end{align*}}

Finally, we show that $\tildeV{2i+1}{m} = (\tildeV{i}{m-1})^{(2)}_{-1}$ in the way discussed above, taking into account that since $2(i+1) \not\in \IS{m}$, there is the one-to-one mapping from $u_{2i+3}^n$ to $\tilde u_{2i+3}^n$:
{\allowdisplaybreaks
\begin{align*}
        \tildeV{2i+1}{m} & (y_1^n, u_1^{2i} | u_{2i+1}, u_{2i+2}) = \\
        =& \max_{\tilde u_{2i+3}^n} \W{}{m-1}(y_{1,o}^n | \tilde u_{1,o}^n) \W{}{m-1}(y_{1,e}^n | \tilde u_{1,e}^n) \\
        =& \begin{aligned}[t]
                & \tildeV{i}{m-1} (y_{1,o}^n, \tilde u_{1,o}^{2i-2} | \tilde u_{2i-1}, \tilde u_{2i+1}) \cdot \\ \cdot & \tildeV{i}{m-1}(y_{1,e}^n, \tilde u_{1,e}^{2i-2} | \tilde u_{2i}, \tilde u_{2i+2})
        \end{aligned} \\
        \stackrel{(f)}{=}& \begin{aligned}[t] 
                & \tildeV{i}{m-1}(y_{1,o}^n, \tilde u_{1,o}^{2i-2} | u_{2i-1} + u_{2i}, u_{2i+1} + u_{2i+2}) \cdot \\ \cdot & \tildeV{i}{m-1}(y_{1,e}^n, \tilde u_{1,e}^{2i-2} | u_{2i}, u_{2i+2})
        \end{aligned} \\
        \stackrel{(g)}{=}& (\tildeV{i}{m-1})^{(2)}_{-1}(y_1^n, \tilde u_1^{2i-2}.u_{2i-1}.u_{2i} | u_{2i+1}, u_{2i+2}).
\end{align*}}

Let us discuss the remaining cases of the lemma. The cases (\ref{item:swp-case}) and (\ref{item:add-case}) can be proved similarly up to replacing \eqref{eq:msg-pretransform} with the corresponding relations. Meanwhile, the cases (\ref{item:ori-case-1})-(\ref{item:ori-case-3}) are already almost proved. Indeed, while proving $\tildeV{2i-1}{m} = (\tildeV{i}{m-1})^{(0)}_{-1}$ we did not use $2(i+1) \in \II{m}$ (we did not use $\tilde u_{2i+1} = u_{2i+1} + u_{2i+2},\ \tilde u_{2i+2} = u_{2i+2}$ from \eqref{eq:msg-pretransform} either; insted, we did a bijective substitution for the "free" variables $\tilde u_{2i+1}$ and $\tilde u_{2i+2}$). On the other hand, for $2(i-1) \in \IS{m}$, our proof of $\tildeV{2i+1}{m} = (\tildeV{i}{m-1})^{(2)}_{-1}$ is still relevant if $u_{2i-1}$ is replaced with $u_{2i-2}$ in $(f)$ and $(g)$. Finally, when $2(i-1) \in \IS{m},\ 2(i+1) \in \II{m}$, one has $\tilde u_{2i-1} = u_{2i-2} + u_{2i},\ \tilde u_{2i+1} = u_{2i+1} + \tilde u_{2i+2}$, so that
{\allowdisplaybreaks
\begin{align*}
        \tildeV{2i}{m}&(y_1^n, u_1^{2i-1}|u_{2i},u_{2i+1}) = \\
        =& \begin{aligned}[t]
                \max_{\tilde u_{2i+2}} & \tildeV{i}{m-1}(y_{1,o}^n, \tilde u_{1,o}^{2i-2} | \tilde u_{2i-1}, \tilde u_{2i+1}) \cdot \\ \cdot & \tildeV{i}{m-1}(y_{1,e}^n, \tilde u_{1,e}^{2i-2} | \tilde u_{2i}, \tilde u_{2i+2})
        \end{aligned} \\
        =& \begin{aligned}[t]
                \max_{\tilde u_{2i+2}} & 
                \tildeV{i}{m-1}(y_{1,o}^n, \tilde u_{1,o}^{2i-2} | u_{2i-2} + u_{2i}, u_{2i+1} + \tilde u_{2i+2}) \cdot \\
        \cdot & \tildeV{i}{m-1}(y_{1,e}^n, \tilde u_{1,e}^{2i-2} | u_{2i}, \tilde u_{2i+2}) \\
        =& (\tildeV{i}{m-1})^{(1)}_{-1}(y_1^n, \tilde u_{1}^{2i-2}.u_{2i-2} | u_{2i},u_{2i+1}).
        \end{aligned}
\end{align*}}

\subsection{Proof of Lemma \ref{lm:approx-llrs}} \label{app:proof-lm-2}

For all $s \in \{-1,0,1\},\ t \in \{1,2\}$ by definition
\begin{equation}
\label{eq:prev-channel}
\begin{aligned}
        V_s^{(t-1)}&(y_1^2,u_1^{t-1} | u_{t}, u_{t+1}) =\max_{u_{t+2} \in \Bin}V_s^{(t)}(y_1^2,u_1^t|u_{t+1},u_{t+2})
\end{aligned}
\end{equation}
proving $L_s^{(t)}(y_1^2,u_1^t) = R_s^{(t-1)}(y_1^2,u_1^{t-1};u_t)$. Next, the proofs of \eqref{eq:llr-oria}, \eqref{eq:llr-orib}, \eqref{eq:llr-oric}, \eqref{eq:llr-oric-right}, \eqref{eq:llr-swpc-right}, \eqref{eq:llr-addc-right} are similar; as an example we show \eqref{eq:llr-orib}:
{\allowdisplaybreaks
\begin{align*}
        R_{-1}^{(0)}&(y_1^2;u_1) = \\
        =& \ln\frac{\max_{u_3,u_4 \in \Bin} V(y_1|u_1, u_3+u_4) V(y_2|0,u_4)} {\max_{u_3,u_4 \in \Bin} V(y_1|\overline u_1, u_3+u_4) V(y_2|1,u_4)} \\
        =& \ln\frac{\max_{u_4 \in \Bin} \left\{ V(y_2|0,u_4) \cdot \max_{x \in \Bin} V(y_1|u_1,x) \right\}}{\max_{u_4 \in \Bin} \left\{ V(y_2|1,u_4) \cdot \max_{x \in \Bin} V(y_1|\overline u_1,x) \right\}} \\
        =& \ln\frac{\left( \max_{x \in \Bin} V(y_1|u_1,x) \right) \cdot \left( \max_{u_4 \in \Bin}V(y_2|0,u_4) \right)}{ \left( \max_{x \in \Bin}V(y_1|\overline u_1,x) \right) \cdot \left( \max_{u_4 \in \Bin} V(y_2|1,u_4) \right)} \\
        =& \ln\frac{\max_{x \in \Bin} V(y_1|u_1,x)}{\max_{x \in \Bin}V(y_1|\overline u_1,x)} + \ln\frac{\max_{u_4 \in \Bin}V(y_2|0,u_4)}{\max_{u_4 \in \Bin} V(y_2|1,u_4)}\\
        =& (-1)^{u_1}L(y_1) + L(y_2) = f_+(L(y_1),L(y_2),u_1)
\end{align*}}
with the used shortcut $\overline u_1 = u_1+1$ and relabeling $u_3 \mapsto x = u_3+u_4$.

Next, we claim that the proofs of \eqref{eq:llr-b} for $s=0$ and $s=1$ are analogous. To explore the case $s=0$, we notice
\begin{equation}
\label{eq:swp-to-ori}
        V_0^{(1)}(y_1^2,u_1 | u_2,u_3) = V_{-1}^{(1)}(y_1^2,u_1 | u_3,u_2).
\end{equation}
Let $\mu(x_1,x_2) \triangleq V_0^{(1)}(y_1^2,u_1|x_1,x_2)$, then 
{\allowdisplaybreaks
\begin{align*}
        R&_0^{(0)}(y_1^2;u_1) \stackrel{(a)}{=} \ln\frac{\max\{\mu(0,0), \mu(0,1)\}} {\max\{\mu(1,0), \mu(1,1)}  \\
        &\stackrel{(b)}{=} \max\{\ln\mu(0,0), \ln\mu(0,1)\} - \max\{\ln\mu(1,0), \ln\mu(1,1)\}\\
        &\stackrel{(c)}{=} \ln\frac{\mu(0,1)}{\mu(1,1)} + \max\left\{ 0,\ln \frac{\mu(0,0)}{\mu(0,1)} \right\} - \max\left\{ 0, \ln\frac{\mu(1,0)}{\mu(1,1)} \right\} \\
        &\stackrel{(d)}{=} \begin{aligned}[t]R_{-1}^{(1)}(y_1^2,u_1;1) +& \max\{0,R_0^{(1)}(y_1^2,u_1;0)\} - \\ -& \max\{0,R_0^{(1)}(y_1^2,u_1;1)\}.
        \end{aligned}
\end{align*}}
The equality $(a)$ is application of \eqref{eq:prev-channel}; $(b)$ stems from $\ln\dfrac{a}{b}=\ln a - \ln b$ and the fact that $\ln$ monotonically increases; $(c)$ follows from $\max\{a,b\} = b + \max\{0,a-b\}$; the transition $(d)$ is obtained by applying \eqref{eq:swp-to-ori} to the first term and the substitution $\ln\dfrac{\mu(b,0)}{\mu(b,1)} = R_0^{(1)}(y_1^2,u_1;b),\ b \in \Bin$. The case $s=1$ can be proved in the same way, considering
\begin{equation*}
        V_1^{(1)}(y_1^2,u_1|u_2,u_3) = V_{-1}^{(1)}(y_1^2,u_1|u_2+u_3,u_3).
\end{equation*}

Finally, we prove \eqref{eq:llr-c-0} restricted to $s=0,\ u_2=0$ (the other cases are similar). With $\eta(x_1,x_2) \triangleq V_{-1}^{(1)}(y_1^2,u_1|x_1,x_2)$,
{\allowdisplaybreaks
\begin{align*}
        R_0^{(1)}&(y_1^2,u_1;0) = \ln\frac{\mu(0,0)}{\mu(0,1)} \stackrel{\eqref{eq:swp-to-ori}}{=} \ln\frac{\eta(0,0)}{\eta(1,0)} \\
        =& \ln\frac{\eta(0,0)}{\eta(0,1)} + \ln\frac{\eta(0,1)}{\eta(1,1)} + \ln\frac{\eta(1,1)}{\eta(1,0)} \\
        \stackrel{(e)}{=}& R_{-1}^{(1)}(y_1^2,u_1;0) - R_{-1}^{(1)}(y_1^2,u_1;1) + L_{-1}^{(1)}(y_1^2,u_1) - \\ -& \max\{0,R_{-1}^{(1)}(y_1^2,u_1;0)\} + \max\{0,R_{-1}^{(1)}(y_1^2,u_1;1)\}.
\end{align*}}
Application of $x - \max\{0,x\} = -\max\{0,-x\}$ terminates the proof. In $(e)$ we used the fact that
\begin{equation*}
        \ln\frac{V(y|0,1)}{V(y|1,1)} = L(y) - \max\{0,R(y;0)\} + \max\{0,R(y;1)\}
\end{equation*}
holds for every DBI channel $V$, in particular for $V \triangleq V_{-1}^{(1)}$.

\bibliographystyle{IEEEtran}
\bibliography{bibliography}{}

\end{document}